\documentclass[12pt,a4paper]{article}
\usepackage{amsfonts}
\usepackage{bm}
\usepackage{amsmath}
\usepackage{epsfig}
\usepackage{breqn}
\usepackage{hyperref}

\newcommand{\bmat}{\left(\begin{array}}
\newcommand{\emat}{\end{array}\right)}

\def\gtrsim{\mathrel{\raise.3ex\hbox{$>$\kern-.75em\lower1ex\hbox{$\sim$}}}}

\def\-{\hphantom{-}}

\def\s2{\frac{1}{\sqrt2}}

\def\mg{m_{3/2}}
\def\mg2{m^2_{3/2}}

\def\Dsl{\,\raise.15ex\hbox{/}\mkern-13.5mu D} 

\def\be{\begin{equation}}
\def\ee{\end{equation}}
\def\bea{\begin{eqnarray}}
\def\eea{\end{eqnarray}}

\newcommand{\nn}{\nonumber}

\begin{document}

\pagestyle{plain}

\makeatletter
\@addtoreset{equation}{section}
\makeatother
\renewcommand{\theequation}{\thesection.\arabic{equation}}
\pagestyle{empty}

\vskip -10mm

\begin{flushright}
\small
YITP-17-18
\end{flushright}

\begin{center}
\ \

\vskip .5cm

\LARGE{\LARGE\bf The Odd story of $\alpha'$-corrections \\[10mm]}
\vskip 0.3cm
\large{Walter H. Baron$^{a,b}$, Jos\'e J. Fern\'andez-Melgarejo$^{c,d}$, \\ Diego Marqu\'es$^e$ and Carmen A. Nu\~nez$^{e,f}$
 \\[6mm]}

{\small\it  $^a$ Instituto de F\'isica La Plata (CONICET-UNLP)\\ }
{\small\it  $^b$ Departamento de F\'isica, Universidad Nacional de La Plata \\ }
{\small\it  $^c$ Yukawa Institute for Theoretical Physics, Kyoto University \\ }
{\small\it  $^d$ Departamento de F\'isica, Universidad de Murcia \\ }
{\small\it  $^e$ Instituto de Astronom\'ia y F\'isica del Espacio (IAFE-CONICET-UBA)\\ }
{\small\it  $^f$ Departamento de F\'isica, FCEyN, Universidad de Buenos Aires \\ [.5 cm]}

{\small \verb"wbaron@fisica.unlp.edu.ar, josejuan@yukawa.kyoto-u.ac.jp, " \\ [.1 cm]}
{\small \verb"diegomarques@iafe.uba.ar, carmen@iafe.uba.ar" \\ [1 cm]}

\small{\bf Abstract} \\[0.5cm]\end{center}

{\small The $\alpha'$-deformed frame-like Double Field Theory (DFT) is a T-duality and gauge invariant extension of DFT in which generalized Green-Schwarz transformations provide a gauge principle that fixes the higher-derivative corrections. It includes all the first order $\alpha'$-corrections of the bosonic and heterotic string low energy effective actions  and of the Hohm-Siegel-Zwiebach $\alpha'$-geometry. Here we gauge this theory  and  parameterize it in terms of a frame, a two-form, a dilaton, gauge vectors and scalar fields. This leads to a unified framework that extends the previous construction by including all duality constrained  interactions in generic (gauged/super)gravity effective field theories in arbitrary number of dimensions,  to first order in $\alpha'$.
}

\newpage
\setcounter{page}{1}
\pagestyle{plain}
\renewcommand{\thefootnote}{\arabic{footnote}}
\setcounter{footnote}{0}

\tableofcontents

\section{Introduction} \label{SEC:Intro}

Double Field Theory (DFT) \cite{Siegel:1993xq}-\cite{Hull:2009mi} reformulates the two-derivative
universal gravitational sector of string theory in such a way that  T-duality symmetry can be anticipated
before dimensional reduction (for reviews see \cite{reviews}). The field has been remarkably active in the last years and much progress has been achieved in several directions: supersymmetrization \cite{SUSY}, extensions that include heterotic \cite{Hohm:2011ex} and type II theories \cite{HetTypeII}, U-duality invariance (Exceptional Field Theory) \cite{EFT}, understanding  duality covariant geometries \cite{GG}, non-geometry \cite{Non-geometry}, finite gauge transformations \cite{Large}, solution generating techniques \cite{Solutions}, etc. The list continues, and most certainly the framework and its applications will further expand in the coming years. In this paper we will focus on two remarkable aspects of DFT: Generalized Scherk-Schwarz compactifications and $\alpha'$-corrections. Let us first discuss them separately.

Generalized Scherk-Schwarz (GSS) compactifications of DFT lead to lower-dimensional gauged supergravities \cite{Aldazabal:2011nj},\cite{Geissbuhler:2013uka}. The information on the compact space (typically a T-fold \cite{T-folds} or a double twisted torus) is encoded in a generalized twist matrix, in terms of which the fluxes can be spelled out. The procedure neatly and efficiently leads to half-maximal gauged supergravities expressed in the embedding tensor formalism \cite{gauged sugra}. The advantage of the approach is that, unlike the standard Scherk-Schwarz (SS) procedure \cite{Scherk:1979zr}, the parent theory is duality invariant and the duality symmetry is preserved all along, without the need to reorganize the degrees of freedom in the effective action into duality multiplets. Moreover, it was shown in \cite{Grana:2012rr} that the result of a GSS compactification of DFT is effectively equivalent to simply gauging the theory and parameterizing the generalized fields in terms of the degrees of freedom of the lower dimensional theory. In addition, the frame or flux formulation of DFT \cite{Siegel:1993xq},\cite{frameflux} allows to define the so-called generalized fluxes, which contain  all the covariant field strengths of the effective theory upon compactification. Then, the compactification procedure is notably simplified  in this formulation, as it offers all the covariant tensors of the theory, even before the effective action is computed. Furthermore, it is possible in this context to relax the strong constraint of DFT in such a way that
all possible duality orbits of gaugings are reached, including the non-geometric ones \cite{Dibitetto:2012rk}. Important work on GSS compactifications can be found in \cite{gSS}.

Another fruitful research direction in DFT points to the understanding of the way in which duality constrains higher-derivative corrections. Since duality covariance must remain unbroken, one seeks for consistent higher-derivative deformations of the gauge transformations, i.e. deformations that  close while keeping the constraints of the theory invariant. The deformations allowed by the generalized metric formulation of DFT turn out to be highly restrictive \cite{Hohm:2016lge}, leaving  the $\alpha'$-geometry of Hohm, Siegel and Zwiebach (HSZ) \cite{Hohm:2013jaa}-\cite{HSZ} as the unique possibility. This theory is  interesting as it is the only known theory that is exactly and manifestly duality invariant and exactly gauge invariant. Instead, deformations that contain the first order $\alpha'$-corrections of the bosonic and heterotic string low energy effective actions are allowed in the frame formalism \cite{Siegel:1993xq},\cite{frameflux}. Actually, a two parameter family of consistent deformations  that  interpolate among the  four-derivative terms of the bosonic and heterotic strings and of the HSZ theory was considered in \cite{Marques:2015vua}. Alternative approaches for the first order $\alpha'$-corrections in this context can be found in \cite{Bedoya:2014pma}.

The aim of this paper is to merge these two frameworks, the GSS compactifications  and the higher-derivative deformations of DFT, into  a gauged $\alpha'$-deformed frame-like DFT.  The outcome of this fusion captures all theories  containing up to four derivative terms of the metric coupled to a two-form, a dilaton, gauge  and scalar fields, constrained by T-duality symmetry. The universe of such theories includes (but is not restricted to) the effective field theories of the closed bosonic string in $26$ dimensions,  heterotic strings in $10$ dimensions (including non Abelian gauge vectors that were not considered in \cite{Marques:2015vua}), and lower dimensional half-maximal gauged supergravities. An interesting aspect of the construction is that the duality group and gauge symmetries completely fix the theory to first order in $\alpha'$ (up to the choice of dimension, interpolating parameters and gauge group), leading to an action that is manifestly invariant under all symmetries, in particular the internal duality group (the subgroup of the original duality group that is preserved in the compactification).  Let us note that computing such theories from standard SS compactifications of, say, the first order heterotic string effective action, would be a highly non-trivial task: not only the degrees of freedom would have to be repackaged into duality multiplets, but this would also require non-covariant field redefinitions. These complications are tractable and easily circumvented in our approach.

Let us add some words on the potential applications and relevance of our results.
Gauged supergravities are the effective lower dimensional field theories that arise from supersymmetry preserving flux compactifications of string theory. Already to lowest order in a derivative expansion, the gaugings lift partially or totally the degeneracy in moduli space, inducing mass terms for scalar and vector fields, and in many cases produce spontaneous supersymmetry breaking. The rich structure of the scalar potential may also induce an effective cosmological constant or determine the dynamics of an inflaton field whose evolution could govern the expansion of the early universe. However, geometric compactifications of two-derivative supergravities are plagued with no-go theorems that prevent many of these nice features from happening, and one is typically led to consider non-geometric compactifications or stringy corrections. Unfortunately, even the leading order corrections to gauged supergravities remain largely unknown. Among our contributions, in this paper we determine the leading order $\alpha'$-corrections to half-maximal gauged supergravities in arbitrary number of dimensions. We especially examine the corrections to the scalar potential and  also perform a preliminary analysis on how the deformations affect the vacuum structure in some simple cases.

The paper is organized as follows. Section \ref{SEC:DFT} reviews some basic aspects of the frame-like formalism of DFT, and the $\alpha'$-deformations considered in \cite{Marques:2015vua}. The generalized Green-Schwarz transformations are displayed, together with the gauge invariant action. The new result contained in this section is  the gauging of the local generalized diffeomorphisms. In Section \ref{SEC:GaugedSugra} we solve the section constraints and present the required parameterizations. This allows to compute the deformed gauge transformations of the components of the generalized fields, and to find the non-covariant field redefinitions to connect with the gauge covariant fields (the frame, two-form, dilaton, gauge and scalar fields). Then, the action is evaluated for these gauge covariant degrees of freedom, and we show how to relate it to the low energy effective actions of the bosonic and heterotic strings and to half-maximal gauged supergravities. Section \ref{SEC:Moduli} is dedicated to explore the corrections to the scalar potential and their effects on the structure of the vacuum. Finally, we conclude in Section \ref{SEC:Conclu}. All the conventions are displayed in Appendix \ref{SEC:Conventions}, which we recommend to visit before reading Section \ref{SEC:GaugedSugra}. The lowest order action, its equations of motion and their relation to covariant first order field redefinitions can be found in Appendix \ref{SEC:Redefs} and Appendix C contains some technical details of the calculations.

\section{$\alpha'$-corrections in Gauged Double Field Theory} \label{SEC:DFT}

We begin by reviewing the (gauged) frame-like formulation of DFT \cite{Siegel:1993xq},\cite{frameflux}. We then show how to deform the theory through the generalized Green-Schwarz transformations of \cite{Marques:2015vua}.

\subsection{Generalized fields, projectors and fluxes}

The frame-like DFT action is invariant under global $G = O(D,D+N|\mathbb{R})$ transformations, local double-Lorentz $H=O(D-1,1|\mathbb{R}) \times O(1, D-1 + N|\mathbb{R})$ transformations, and infinitesimal
generalized diffeomorphisms generated by a generalized Lie derivative $\widehat {\cal L}$. A constant symmetric and invertible $G$-invariant metric $\eta_{M N}$
raises and lowers the indices that are rotated by $G$ (which we label $M,N,\dots = 1,\dots,2D +N$). In addition, there are two constant symmetric and invertible $H$-invariant
metrics $\eta_{A B}$ and ${\cal H}_{A B}$. The former is used to raise
and lower the indices that are rotated by $H$ (which we label $A,B,\dots, K = 1,\dots,2D+N$), and the
latter is
constrained to satisfy
\be
{\cal H}_{A}{}^C {\cal H}_C{}^B = \delta_A^B \ . \label{flatconstraint}
\ee
The three metrics $\eta_{M N}$, $\eta_{A B}$ and ${\cal H}_{A B}$ are invariant under the action of $\widehat{\cal L}$, $G$ and $H$.

The theory is defined on a double space, in which derivatives $\partial_M$ transform in the fundamental representation of $G$.
However, a strong constraint
\be
\partial_M \partial^M \dots = 0 \ , \ \ \ \ \ \partial_M \dots \ \partial^M \dots = 0 \ , \label{StrongConstraint}
\ee
restricts  the coordinate dependence of fields and gauge parameters,
the dots representing arbitrary products of them. The strong constraint is duality invariant, and has the interesting feature that even if its solutions spontaneously break the symmetry, there is no need to specify a particular solution so duality invariance can be maintained. The generalized Lie derivative is generated by an infinitesimal generalized parameter $\xi^M$ that transforms in the fundamental representation of $G$, and $H$-transformations are generated by an infinitesimal parameter $\Lambda_{A}{}^B$ which is constrained by the fact that $\eta_{A B}$ and ${\cal H}_{A B}$ must be $H$-invariant
\be
\delta_\Lambda \eta_{A B} = 2 \Lambda_{(A B)} = 0 \ , \ \ \ \ \ \delta_\Lambda {\cal H}_{A B} = 2 {\cal H}_{C (A} \Lambda^C{}_{B)} = 0 \ . \label{constraintsLambda}
\ee

The fields of the theory are a generalized frame $E_M{}^A$ and a generalized dilaton $d$. The generalized frame is constrained to relate the metric $\eta_{A B}$ with $\eta_{M N}$, and allows to define a generalized metric  ${\cal H}_{M N}$ from ${\cal H}_{A B}$
 \be
\eta_{M N} = E_M{}^A \eta_{A B} E_N{}^B     \ , \ \ \ \ \ \      {\cal H}_{M N} = E_M{}^A {\cal H}_{A B} E_N{}^B \ . \label{flattocurve}
\ee
In general $E_M{}^A$ converts $G$-indices into $H$-indices and vice versa. As a result of (\ref{flatconstraint}), the generalized metric is constrained to be $G$-valued
\be
{\cal H}_{M}{}^P {\cal H}_{P}{}^N = \delta_{M}^N \ . \label{curveconstraint}
\ee

Since the metrics ${\cal H}_{A B}$ and ${\cal H}_{M N}$ are constrained by (\ref{flatconstraint}) and (\ref{curveconstraint}), one can define the following projectors
\be
P = \frac 1 2 \left( \eta - {\cal H}\right) \ , \ \ \ \ \ \bar P = \frac 1 2 \left( \eta + {\cal H}\right) \ , \label{projectors}
\ee
which satisfy the following identities
\be
P^2 = P \ , \ \ \ \ \ \bar P^2 = \bar P \ , \ \ \ \ \ P \bar P = \bar P P = 0 \ . \label{projectoridentities}
\ee
Another useful identity is
\be
P_{M}{}^N E_N{}^A = E_M{}^B P_B{}^A \ , \ \ \ \ \ \ \bar P_{M}{}^N E_N{}^A = E_M{}^B \bar P_B{}^A \ .
\ee
We will use the barred-index notation to denote projections
\be
P_M{}^N V_N = V_{\underline{M}} \ , \ \ \ \ \ \bar P_M{}^N V_N = V_{\overline{M}} \ ,
\ee
and  the following convention for (anti-)symmetrization of
barred-indices
\be
V_{(\underline{M}} W_{\overline{N})} = \frac 1 2 \left( V_{\underline{M}} W_{\overline{N}} + V_{\underline{N}} W_{\overline{M}}\right) \ , \ \ \ \ \ V_{[\underline{M}} W_{\overline{N}]} = \frac 1 2 \left( V_{\underline{M}} W_{\overline{N}} - V_{\underline{N}} W_{\overline{M}}\right)\ ,
\ee
i.e.,  only the indices are exchanged and not the bars.

DFT admits deformations in terms of so-called fluxes or gaugings $f_{M N P}$ \cite{Hohm:2011ex}, a set of constants that satisfy linear and quadratic constraints
\be
f_{M N P} = f_{[M N P]} \ , \ \ \ \ f_{[M N}{}^R f_{P] R}{}^Q = 0 \ . \label{Constraintsfs}
\ee
In the presence of these deformations,  consistency of the theory requires, apart form the strong constraint (\ref{StrongConstraint}), the following additional constraint to further restrict the coordinate dependence of fields and gauge parameters
\be
f_{M N}{}^P \, \partial_P \dots = 0 \ . \label{fderivative}
\ee
This prevents the fields and gauge parameters to depend on coordinates oriented along the directions that are gauged. The gaugings explicitly break the $G$-invariance, unless they are allowed to transform as spurionic $G$-tensors.

Important objects in the frame-like or flux-formulation of DFT are the generalized fluxes
\bea
{\cal F}_{A B C} &=& 3 \partial_{[A} E^N{}_B E^P{}_{C]} \eta_{N P} + f_{M N P} E^M{}_A E^N{}_B E^P{}_C
\ ,
\nn \\
{\cal F}_A &=& 2 \partial_A d - \partial_{B}E_{M A} E^{M B}
\ ,
\eea
and the following projections take a predominant role in the $\alpha'$-deformed theory
\bea
{\cal F}^{(-)}_{M A B} = {\cal F}_{\overline M \underline A \underline B}=\bar P_M{}^NE_N{}^C{\cal F}_{CDE}P_A{}^DP_B{}^E\ , \nn  \\
{\cal F}^{(+)}_{M A B} = {\cal F}_{\underline M \overline A \overline B} =P_M{}^NE_N{}^C{\cal F}_{CDE}\bar P_A{}^D\bar P_B{}^E \ .
\eea

\subsection{Generalized Green-Schwarz transformations}

The generalized dilaton and frame transform under generalized diffeomorphisms and $H$-transformations as follows
\bea
\delta d &=& \xi^P \partial_P d - \frac 1 2 \partial_P \xi^P  \ \ \ \ \Leftrightarrow \ \ \ \ \delta e^{-2d} = \partial_P \left(\xi^P e^{-2 d}\right) \, , \label{vard}\\
\delta E_M{}^A &=& \widehat {\cal L}_\xi E_M{}^A + \delta_\Lambda E_M{}^A  + \widetilde \delta_\Lambda E_M{}^A \ , \label{varE}
\eea
where
the generalized Lie derivative governing infinitesimal generalized diffeomorphisms
is given by
\be
\widehat {\cal L}_\xi E_M{}^A = \xi^P \partial_P E_M{}^A + \left(\partial_M \xi^P - \partial^P \xi_M\right) E_P{}^A + f_{M P}{}^Q \xi^P E_Q{}^A \ ,
\ee
and $H$-transformations split into double-Lorentz transformations
\be
\delta_\Lambda E_M{}^A = E_M{}^B \Lambda_B{}^A \ , \label{DeltaLambdaE}
\ee
and a first order in $\alpha'$ generalized Green-Schwarz transformation \cite{Marques:2015vua}
\be
\widetilde \delta_\Lambda E_M{}^A =  \left( a\, \partial_{[\underline M} \Lambda_{C}{}^{B} \, {\cal F}^{(-)}_{\overline N] B}{}^{C} -  b\, \partial_{[\overline M} \Lambda_{C}{}^{B} \, {\cal F}^{(+)}_{\underline N] B}{}^{C}\right) E^{N A} \ ,\label{genGSgenframe}
\ee
where the parameters $(a,\, b)$ are both of ${\cal O}(\alpha')$. Note that the $P$ and $\bar P$ projections involved in both terms are opposite to each other, so $a$ and $b$ interpolate between generalized Green-Schwarz transformations with respect to the two different factors of the $H$-group.  The fact that there are two free parameters $(a,b)$ implies that we will end with a two-parameter family of theories.  Choosing an appropriate parameterization, it was shown in \cite{Marques:2015vua} that the cases $ (a,b) = (-\alpha', 0)$ and  $(a,b) = (-\alpha', -\alpha')$ correspond to the heterotic and bosonic strings respectively. These cases will be discussed in more detail in the forthcoming sections. The case $(a,b) = (-\alpha', \alpha')$ reproduces the HSZ theory which contains no Riemann squared terms and  the  first order contributions are
given only by Chern-Simons corrections to the curvature of the two-form.

For the generalized metric these transformations imply
\be
\delta {\cal H}_{M N} = \widehat {\cal L}_\xi {\cal H}_{M N} + \widetilde \delta_\Lambda {\cal H}_{M N} \ , \label{vargenmet}
\ee
with
\be
\widehat {\cal L}_\xi {\cal H}_{M N} = \xi^P \partial_P {\cal H}_{M N}\ + 2 \left(\partial_{(M} \xi^P - \partial^P \xi_{(M}\right) {\cal H}_{N) P} - 2 f_{P(M}{}^Q {\cal H}_{N)Q} \xi^P \ ,
\ee
and
\be
\widetilde \delta_\Lambda {\cal H}_{M N} = 2 a\, \partial_{(\underline M} \Lambda_{A}{}^{B} \, {\cal F}^{(-)}_{\overline N) B}{}^{A} + 2 b\, \partial_{(\overline M} \Lambda_{A}{}^{B} \, {\cal F}^{(+)}_{\underline N) B}{}^{A} \ . \label{genGSgenmetric}
\ee

Regarding the generalized fluxes, to lowest order in $\alpha'$ they transform as
\be
\delta {\cal F}_{A B C} =  \xi^P \partial_P {\cal F}_{A B C} - 3 \left( \partial_{[A}\Lambda_{B C]} + \Lambda_{[A}{}^D {\cal F}_{B C] D}\right) \ ,
\ee
which implies that the projected generalized fluxes transform as connections to lowest order
\bea
\delta {\cal F}^{(-)}_{M A}{}^B &=& \widehat {\cal L}_\xi  {\cal F}^{(-)}_{M A}{}^B - \partial_{\overline{M}} \Lambda_{\underline{A}}{}^{\underline{B}} + {\cal F}^{(-)}_{M A}{}^C \Lambda_C{}^B - \Lambda_A{}^C {\cal F}^{(-)}_{M C}{}^B \, ,\nn \\
\delta {\cal F}^{(+)}_{M A}{}^B &=& \widehat {\cal L}_\xi  {\cal F}^{(+)}_{M A}{}^B - \partial_{\underline{M}} \Lambda_{\overline{A}}{}^{\overline{B}} + {\cal F}^{(+)}_{M A}{}^C \Lambda_C{}^B - \Lambda_A{}^C {\cal F}^{(+)}_{M C}{}^B \ , \label{varprojectedF}
\eea
with
\be
\widehat {\cal L}_\xi {\cal F}^{(\pm)}_{M A}{}^B = \xi^P \partial_P {\cal F}^{(\pm)}_{M A}{}^B + \left(\partial_M \xi^P - \partial^P \xi_M \right) {\cal F}^{(\pm)}_{P A}{}^B + f_{M P}{}^Q \xi^P {\cal F}^{(\pm)}_{Q A}{}^B\ .
\ee

The above transformations preserve the constraints on the generalized fields (\ref{flatconstraint})-(\ref{curveconstraint}), and also close to first order in $\alpha'$
\be
\left[ \delta_{(\xi_1 \, , \, \Lambda_1)} , \delta_{(\xi_2 \, , \, \Lambda_2)} \right] = \delta_{(\xi_{21} \, , \, \Lambda_{21})} \ ,
\ee
where the ``brackets'' are given by
\bea
\xi_{12}^M \!&=&\! \left[\xi_1 \, , \, \xi_2 \right]^M_{(C_f)} - \frac a 2 \Lambda_{[1\, \underline A}{}^{\underline B} \partial^M \Lambda_{2]\, \underline B}{}^{\underline A} + \frac b 2 \Lambda_{[1\, \overline A}{}^{\overline B} \partial^M \Lambda_{2]\, \overline B}{}^{\overline A} \, , \label{alphaprimebracket}\\
\Lambda_{12\, A B} \!&=&\! 2 \xi_{[1}^P \partial_P \Lambda_{2] \, A B} - 2 \Lambda_{[1\, A}{}^C \Lambda_{2]\, C B}  +\, a \, \partial_{[\overline A}\Lambda_1^{\underline C \underline D} \partial_{\overline B]} \Lambda_{2 \underline D \underline C} +\, a \, \partial_{[\underline A}\Lambda_1^{\underline C \underline D} \partial_{\underline B]} \Lambda_{2 \underline D \underline C} \nn\\
\!&&\!- \, b \, \partial_{[\underline A}\Lambda_1^{\overline C \overline D} \partial_{\underline B]} \Lambda_{2 \overline D \overline C} - \, b \, \partial_{[\overline A}\Lambda_1^{\overline C \overline D} \partial_{\overline B]} \Lambda_{2 \overline D \overline C}\ ,
\eea
and the $C_f$-bracket is defined as \cite{Hohm:2011ex}
\be
\left[\xi_1 \, , \, \xi_2 \right]^M_{(C_f)} = 2 \xi_{[1}^P \partial_P \xi_{2]}^M  - \xi_{[1}^P \partial^M \xi_{2]P}  + f_{P Q}{}^M \xi_1^P \xi_2^Q\ .
\ee

\subsection{Gauge invariant action}
We now have all the ingredients to construct a gauge-invariant action to first order in $\alpha'$. It can be written as
\be
S = \int dX e^{-2 d}\left({\cal R} + a \, {\cal R}^{(-)} + b \, {\cal R}^{(+)} \right) \ , \label{ActionAlphaPrime}
\ee
where $\cal R$ is of course defined in the same way as in the zeroth order DFT action \cite{Hull:2009mi}
\bea
{\cal R} &=& 4 {\cal H}^{M N} \partial_{M N}{d} - \partial_{M N}{{\cal H}^{M N}} - 4 {\cal H}^{M N} \partial_{M}{ d } \partial_{N}{ d } + 4 \partial_{M}{{\cal H}^{M N} } \partial_{N}{ d } \nn \\
&& + \frac 1 8  {\cal H}^{M N} \partial_{M}{ {\cal H}^{K L} } \partial_{N}{ {\cal H}_{K L} } - \frac 1 2 {\cal H}^{M N} \partial_{M}{ {\cal H}^{K L} } \partial_{K}{ {\cal H}_{N L} } \ . \label{DFTaction}
\eea
Alternatively, the generalized Ricci scalar can also be written in terms of generalized fluxes \cite{Geissbuhler:2013uka}
\bea
{\cal R} &=& (2 \partial_A {\cal F}_B - {\cal F}_A {\cal F}_B) ({\cal H}^{A B} - \eta^{A B}) - \frac 1 4 {\cal F}_{A C}{}^D {\cal F}_{B D}{}^C {\cal H}^{A B}
\nn \\ &&-\, \frac 1 {12} {\cal F}_{A B}{}^C {\cal F}_{D E}{}^F {\cal H}^{A D} {\cal H}^{B E} {\cal H}_{C F} - \frac 1 6 {\cal F}_{A B C} {\cal F}^{A B C} \ . \label{DFTactionfluxes}
\eea

While $\cal R$ is a scalar under generalized diffeomorphisms, it fails to be gauge invariant under the generalized Green-Schwarz transformations (\ref{genGSgenmetric}). Then, additional contributions to the Lagrangian (which must also transform as scalars under generalized diffeomorphisms) need to be considered to compensate for this failure. The generalized Green-Schwarz transformations then constitute a gauge principle that requires and fixes the form of the $\alpha'$-corrections. The required additional first-order corrections are given by
\begingroup\makeatletter\def\f@size{11}\check@mathfonts
\begin{dmath*}[compact]
{\cal R}^{(-)} = {\partial}_{A}{{\partial}_{B}{{\cal F}_{C D\,  E}}\, }\,  {\cal F}_{F G H} \left( \vphantom{\frac 1 2} \right. - {P}^{C F} {P}^{D\,  G} {\bar P}^{A E} {\bar P}^{B H} - {P}^{C F} {P}^{D\,  G} {\bar P}^{A H} {\bar P}^{B E}\left. \vphantom{\frac 1 2} \right) + {\partial}_{A}{{\cal F}_{B C D\, }}\,  {\partial}_{E}{{\cal F}_{F G H}}\,  \left( \vphantom{\frac 1 2} \right.\frac{1}{2}\, {P}^{A E} {P}^{B F} {P}^{C G} {\bar P}^{D\,  H} - {P}^{B F} {P}^{C G} {\bar P}^{A D\, } {\bar P}^{E H} - \frac{1}{2}\, {P}^{B F} {P}^{C G} {\bar P}^{A E} {\bar P}^{D\,  H}\left. \vphantom{\frac 1 2} \right) + (2\, {\partial}_{A}{{\cal F}_{B}} - {\cal F}_A {\cal F}_B)\,  {\cal F}_{C D\,  E} {\cal F}_{F G H} {P}^{C F} {P}^{D\,  G} {\bar P}^{A E} {\bar P}^{B H} + {\partial}_{A}{{\cal F}_{B C D\, }}\,  {\cal F}_{F G H} {\cal F}_{E} \left( \vphantom{\frac 1 2} \right.2\, {P}^{B F} {P}^{C G} {\bar P}^{A D\, } {\bar P}^{E H} + 2\, {P}^{B F} {P}^{C G} {\bar P}^{A H} {\bar P}^{D\,  E}\left. \vphantom{\frac 1 2} \right) + {\partial}_{A}{{\cal F}_{B C D\, }}\,  {\cal F}_{E F G} {\cal F}_{H I J} \left( \vphantom{\frac 1 2} \right. - {P}^{A E} {P}^{B H} {P}^{C I} {\bar P}^{D\,  F} {\bar P}^{G J} - 4\, {P}^{B E} {P}^{C H} {P}^{F I} {\bar P}^{A G} {\bar P}^{D\,  J} + {P}^{B E} {P}^{C F} {\bar P}^{A H} {\bar P}^{D\,  I} {\bar P}^{G J}\left. \vphantom{\frac 1 2} \right)  + {\cal F}_{A B C} {\cal F}_{D\,  E F} {\cal F}_{G H I} {\cal F}_{J K L} \left( \vphantom{\frac 1 2} \right.{P}^{A D\, } {P}^{B G} {P}^{E J} {P}^{H K} {\bar P}^{C L} {\bar P}^{F I} - {P}^{A D\, } {P}^{B G} {P}^{E J} {P}^{H K} {\bar P}^{C F} {\bar P}^{I L} + {P}^{A D\, } {P}^{B E} {P}^{G J} {\bar P}^{C H} {\bar P}^{F K} {\bar P}^{I L} + \frac{4}{3}\, {P}^{A D\, } {P}^{B G} {P}^{E H} {\bar P}^{C J} {\bar P}^{F K} {\bar P}^{I L}\left. \vphantom{\frac 1 2} \right) \ ,
\end{dmath*}
\endgroup
\noindent and ${\cal R}^{(+)}$ can be easily obtained from this through the substitution ${\cal R}^{(+)} = {\cal R}^{(-)}[P \leftrightarrow \bar P]$. The ungauged limit of this action reduces to that in \cite{Marques:2015vua}.

The three contributions to the Lagrangian ${\cal R}$ and ${\cal R}^{(\pm)}$ are generalized diffeomorphism scalars (modulo the constraints (\ref{StrongConstraint}) and (\ref{fderivative})), and the full Lagrangian  is $H$-invariant to first order in $\alpha'$
\be
\delta \left({\cal R}+ a {\cal R}^{(-)} + b {\cal R}^{(+)} \right) = \widehat {\cal L}_\xi \left({\cal R}+ a {\cal R}^{(-)} + b {\cal R}^{(+)} \right)  \ .
\ee
In fact, one can show that the anomalous Lorentz behaviour $\widetilde \delta_\Lambda {\cal R}$ is exactly cancelled by $\delta_\Lambda \left(a \, {\cal R}^{(-)} + b \, {\cal R}^{(+)}\right)$. Notice also that $\widetilde \delta_\Lambda \left(a \, {\cal R}^{(-)} + b \, {\cal R}^{(+)}\right)$ is of higher order, and must then not be considered in this computation. We conclude that the action (\ref{ActionAlphaPrime}) is exactly invariant under $G$ and $\widehat {\cal L}$ symmetries, and $H$-invariant to ${\cal O}(\alpha')$.

 \section{$\alpha'$-corrections in Gauged Supergravity} \label{SEC:GaugedSugra}

So far our construction has been general: we have assumed neither a parameterization of the generalized fields nor a solution to the strong constraint (\ref{StrongConstraint}). Here we give the parameterizations required to make contact with  theories of gravity coupled to a two-form, a dilaton, gauge vectors and scalar fields. We have chosen the duality group and its pseudo-compact subgroup to be
\bea
G &=& O(D,D+N|\mathbb{R}) \ , \\
H &=& O(D-1,1|\mathbb{R}) \times O(1, D-1 + N|\mathbb{R})\ .
\eea
We now assume a splitting of the form $D = n + d$, such that the $G$-indices split as $V^M = (V_\mu,\, V^\mu,\, V^m)$ and the $H$-indices split as $V^A = (V_a,\, V^a,\, V^\alpha)$, where $\mu,a = 1,\dots,n$ (external) and $m,\alpha=1,\dots,2d + N$ (internal). This splitting spontaneously breaks $G$ and $H$ into external and internal parts
\be
G \to G_e \times G_i\ , \ \ \ \ \
H \to H_e \times H_i\  ,
\ee
where
\bea
G_e &=& O(n,n|\mathbb{R}) \ , \\
G_i &=& O(d,d+N|\mathbb{R}) \ , \\
H_e &=& O(n-1,1|\mathbb{R}) \times O(1,n-1|\mathbb{R}) \ , \\
H_i &=& O(d|\mathbb{R}) \times O(d + N|\mathbb{R})\ .
\eea
Then, the $G$-vector $V^M$ contains a $G_e$-vector $(V_\mu,\, V^\mu)$ and a $G_i$-vector $V^m$, and the $H$-vector $V^A$ contains a $H_e$-vector $(V_a,\, V^a)$ and a $H_i$-vector $V^\alpha$.

Under this decomposition, the degrees of freedom can be parameterized as follows
\bea
{\rm dim}(G/H) = D(D+N) &=& \frac {n (n+1)} 2 + \frac {n (n-1)} 2 + ~ n (2d + N) ~ + ~ d(d+N)  \nn\\
&& \ \ \ \ \widetilde g_{\mu \nu} \ \ \ \ \ \ \ \ \ \ \  \widetilde B_{\mu \nu} \ \ \ \ \ \ \ \ \ \ \ \  \ \widetilde A_\mu{}^m \ \ \ \ \ \ \ \ \ \ \ \ \widetilde \Phi_m{}^\alpha\
\eea
where $\widetilde g_{\mu \nu}$ is symmetric and invertible, $\widetilde B_{\mu \nu}$ is antisymmetric, and $\widetilde \Phi_m{}^\alpha$ parameterizes the coset $G_i/H_i$.

\subsection{Parameterization and choice of section}

The matrices $\eta_{A B}$, ${\cal H}_{A B}$ and $\eta_{M N}$ are taken to be
 \be
\eta_{A B} = \left(\begin{matrix} 0 & \delta^a_b & 0 \\ \delta^b_a & 0 & 0 \\ 0 & 0 & \kappa_{\alpha \beta}\end{matrix}\right) \ , \ \ \ \ {\cal H}_{A B} = \left(\begin{matrix} g^{a b} & 0 & 0 \\ 0 & g_{a b} & 0 \\ 0 & 0 & M_{\alpha \beta}\end{matrix}\right) \ , \ \ \ \ \eta_{M N} = \left(\begin{matrix} 0 & \delta^\mu_\nu & 0 \\ \delta^\nu_\mu & 0 & 0 \\ 0 & 0 & \kappa_{m n}\end{matrix}\right) \ , \label{ParamMetrics}
\ee
where $g_{a b}$ is the flat Minkowski metric in the external space, $\kappa_{\alpha \beta}$ and $M_{\alpha \beta}$ are the two $H_i$-invariant matrices, and $\kappa_{m n}$ is the $G_i$ invariant metric. Internal $G_i$ and $H_i$-indices are raised and lowered with $\kappa_{m n}$ and $\kappa_{\alpha \beta}$  respectively.
The gaugings are chosen to be non-vanishing only in the internal directions
\be
f_{M N P} = \left\{ \begin{matrix} f_{m n p} \ \ \ \ {\rm if}\  (M,N,P) = (m,n,p) \\
                                    0 \ \ \ \ \ \  {\rm otherwise}\ \ \ \ \ \ \ \ \ \ \ \ \ \ \ \ \end{matrix}\right. \ .
\ee
For the $f_{mnp}$ gaugings, the linear and quadratic constraints (\ref{Constraintsfs}) straightforwardly translate into
 \be
f_{mnp} = f_{[mnp]} \ , \ \ \ \ f_{[mn}{}^r f_{p]r}{}^q = 0 \ . \label{Constraintsfsmall}
\ee
A natural solution to the constraints (\ref{StrongConstraint}) and (\ref{fderivative}) is
\be
\partial_M = (\tilde \partial^{\mu} , \partial_\mu , \partial_m) = \left(0 , \partial_\mu , 0\right) \ , \label{SolSC}
\ee
so the fields will only depend on the $X^\mu$ coordinates of the $n$-dimensional external space.

The generalized frame is parameterized as follows
\be
E_M{}^A = \left(\begin{matrix} \widetilde e_a{}^\mu & 0 & 0 \\ -\widetilde e_a{}^\rho \widetilde C_{\rho \mu} & \widetilde e_\mu{}^a & \widetilde A_\mu{}^p \widetilde \Phi_p{}^\alpha \\ - \widetilde e_a{}^\rho \widetilde A_{\rho m} & 0 & \widetilde \Phi_m{}^\alpha\end{matrix}\right) \ .
\ee
Here $\widetilde e_\mu{}^a$ is the frame for $\widetilde g_{\mu \nu} = \widetilde e_\mu{}^a g_{a b} \widetilde e_\nu{}^b$, and $\widetilde e_a{}^\mu = g_{a b} \widetilde g^{\mu \nu} \widetilde e_\nu{}^b$ is the inverse frame. They satisfy the identities $\widetilde e_\mu{}^a \widetilde e_a{}^\nu = \delta_\mu^\nu$ and $\widetilde e_a{}^\mu \widetilde e_\mu{}^b = \delta^b_a$. Also, we have defined
\be
\widetilde C_{\mu \nu} = \widetilde B_{\mu \nu} + \frac 1 2  \widetilde A_\mu{}^m \widetilde A_{\nu m} \ .
\ee
The parameterization of the generalized metric that follows from the above choices is
\be
{\cal H}_{M N} = \left(\begin{matrix} \widetilde g^{\mu \nu} & - \widetilde g^{\mu \rho} \widetilde C_{\rho \nu} & - \widetilde g^{\mu \rho} \widetilde A_{\rho n} \\
                 - \widetilde g^{\nu \rho} \widetilde C_{\rho \mu} & \widetilde g_{\mu \nu} + \widetilde C_{\rho \mu} \widetilde C_{\sigma \nu} \widetilde g^{\rho \sigma} + \widetilde A_\mu{}^p \widetilde M_{pq} \widetilde A_{\nu}{}^q &
                 \widetilde C_{\rho \mu} \widetilde g^{\rho \sigma} \widetilde A_{\sigma n} + \widetilde A_{\mu}{}^p \widetilde M_{p n} \\
                 - \widetilde g^{\nu \rho} \widetilde A_{\rho m} & \widetilde C_{\rho \nu} \widetilde g^{\rho \sigma} \widetilde A_{\sigma m} + \widetilde A_\nu{}^p \widetilde M_{m p} & \widetilde M_{m n} + \widetilde A_{\rho m} \widetilde g^{\rho \sigma} \widetilde A_{\sigma n}\end{matrix}\right) \ ,
\ee
where we have defined the internal (scalar) matrix $\widetilde M_{m n} = \widetilde \Phi_m{}^\alpha M_{\alpha \beta} \widetilde \Phi_n{}^\beta$, which is symmetric and $G_i$-valued $\widetilde M_m{}^p \widetilde M_p{}^n = \delta^n_m$. We also define the inverse internal frame as $\widetilde \Phi_\alpha{}^m = \kappa_{\alpha \beta} \kappa^{m n} \widetilde \Phi_n{}^\beta$, which satisfies $\widetilde \Phi_m{}^\alpha \widetilde \Phi_\alpha{}^n = \delta^n_m$ and $\widetilde \Phi_\alpha{}^m \widetilde \Phi_m{}^\beta = \delta^\beta_\alpha$.

The parameterization of the generalized dilaton  is given by
\be
e^{- 2 d} = \sqrt {- \widetilde g} e^{-2 \widetilde \phi} \ .
\ee

We finally turn to the parameterization of the gauge parameters $\xi^M$ and $\Lambda_{A B}$
\be
\xi^M = \left( \xi_\mu , \xi^\mu , \lambda^m \right) \ , \ \ \ \Lambda_{A B} = \left(\begin{matrix} 0 & \Lambda^a{}_b & 0 \\ \Lambda_a{}^b & 0 & 0 \\ 0 & 0 & \Lambda_{\alpha \beta}\end{matrix}\right) \ .
\ee
Note that the choice of parameterization of the generalized frame assumes a gauge-fixing of the external double-Lorentz transformations to the diagonal part corresponding to the standard single Lorentz transformations parameterized here by $\Lambda_{a b}$. On the other hand, $\Lambda_{\alpha \beta}$ are the infinitesimal parameters that generate $H_i$-transformations.

We have put a tilde on top of all the fields because, due to the generalized Green-Schwarz transformation, these components receive $\alpha'$-correction corrections in their gauge transformations, and then they are related to the corresponding gauge covariant fields in supergravity through first order in $\alpha'$ Lorentz non-covariant field redefinitions. Of course, to lowest order, these fields are precisely the corresponding fields in gauged supergravity, which we will denote without tildes. We then expect an expansion of the form
\bea
 &&\widetilde e_\mu{}^a = e_\mu{}^a + {\cal O}(\alpha') \ , \ \ \ \widetilde B_{\mu \nu} = B_{\mu \nu} + {\cal O}(\alpha') \ , \ \ \ \widetilde \phi = \phi  + {\cal O}(\alpha') \nn \\
 && \widetilde A_\mu{}^m = A_\mu{}^m  + {\cal O}(\alpha') \ , \ \ \ \widetilde \Phi_m{}^\alpha = \Phi_m{}^\alpha  + {\cal O}(\alpha') \ .
\eea
We now introduce expressions for the generalized fluxes. To lighten the notation, we remove the tildes from the fields, so strictly speaking the following identities only hold to lowest order. The exact expressions are simply recovered by reinstalling tildes on all fields
\bea
{\cal F}_{a b c} &=& - e_a{}^\mu e_b{}^\nu e_c{}^\rho H_{\mu \nu \rho} \label{Ffrom1}
\ ,\\
{\cal F}_{a b}{}^c &=& - 2 e_{[a}{}^\mu \omega_{\mu b]}{}^c
\ ,\\
{\cal F}_{a b \alpha} &=& - \Phi_\alpha{}^m e_a{}^\mu e_b{}^\nu F_{\mu \nu m} \label{Ffrom}
\ ,\\
{\cal F}_{a \alpha \beta} &=& - e_a{}^\mu \omega_{\mu \alpha \beta}
\ ,\\
{\cal F}_{\alpha \beta \gamma} &=& \Phi_\alpha{}^m \Phi_\beta{}^n \Phi_\gamma{}^p f_{m n p} \label{Fto}
\ ,\\
{\cal F}_a &=& 2 e_a{}^\mu \partial_\mu \phi + e_b{}^\mu \omega_{\mu a}{}^b
\ .
\label{Fto2}
\eea
All these quantities are defined in Appendix \ref{SEC:Conventions}.

Given that $\kappa_{\alpha \beta}$ raises and lowers the $H_i$-indices, and that $M_\alpha{}^\gamma M_\gamma{}^\beta = \delta^\beta_\alpha$, it follows that one can define projectors as before
\be
P_{\alpha \beta} = \frac 1 2 \left(\kappa_{\alpha \beta} - M_{\alpha \beta}\right) \ , \ \ \ \ \bar P_{\alpha \beta} = \frac 1 2 \left(\kappa_{\alpha \beta} + M_{\alpha \beta}\right) \ , \label{internalPs1}
\ee
in analogy with (\ref{projectors}), that satisfy relations equivalent to (\ref{projectoridentities}). It is also convenient to define projected $H_i$-indices $\alpha = \underline{\alpha} + \overline{\alpha}$ as follows
\be
P_\alpha{}^\beta V_\beta = V_{\underline{\alpha}} \ , \ \ \ \ \bar P_\alpha{}^\beta V_\beta = V_{\overline{\alpha}} \ . \label{internalPs2}
\ee
\subsection{Gauge transformations and field redefinitions}

Let us now explore how the tilded fields $\widetilde g_{\mu \nu}$, $\widetilde B_{\mu \nu}$, $\widetilde \phi$, $\widetilde A_\mu{}^m$ and $\widetilde M_{m n}$ transform under (\ref{vard})-(\ref{varE}). Implementing the parameterization of the previous subsection we find
\bea
\delta \widetilde g_{\mu \nu}
&=&
L_\xi \widetilde g_{\mu \nu} + \frac 1 2 \left(a \omega^{(-)}_{(\mu}{}^{\alpha \beta } + b \omega^{(+)}_{(\mu}{}^{\alpha \beta}\right) \partial_{\nu)}\Lambda_{\alpha \beta} + \frac 1 2 \left(a \omega^{(-)}_{(\mu}{}^{a b} + b \omega^{(+)}_{(\mu}{}^{a b} \right) \partial_{\nu)}\Lambda_{a b} \ , \ \ \ \ \ \
 \\
\delta \widetilde A_\mu{}^m &=& L_\xi \widetilde A_\mu{}^m + \partial_\mu \lambda^m + f_{p q}{}^m \lambda^p \widetilde A_\mu{}^q - \frac 1 2 \left(a {\cal F}^{\underline{\alpha} \underline{\beta} \overline{\gamma}} - b {\cal F}^{\overline{\alpha} \overline{\beta} \underline{\gamma}}\right) \Phi_\gamma{}^m \partial_\mu \Lambda_{\alpha \beta}
 \nn \\
 && + \frac 1 4 \left(a {\cal F}^{a b \overline{\alpha}} + b {\cal F}^{a b \underline{\alpha}}\right) \Phi_\alpha{}^m \partial_\mu \Lambda_{a b}
\ , \\
\delta \widetilde B_{\mu \nu} &=& L_\xi \widetilde B_{\mu \nu} + 2 \partial_{[\mu} \xi_{\nu]} + \widetilde A_{[\mu}{}^m \partial_{\nu]}\lambda_m
       - \frac 1 2 \left(a \omega^{(-)}_{[\mu}{}^{\alpha \beta} - b \omega^{(+)}_{[\mu}{}^{\alpha \beta}\right) \partial_{\nu]}\Lambda_{\alpha \beta}
        \nn \\
       && - \frac 1 2 \left(a \omega_{[\mu}^{(-)ab} - b \omega_{[\mu}^{(+)ab}\right) \partial_{\nu]}\Lambda_{ab} + \frac 1 2 \left(a {\cal F}^{\underline{\alpha} \underline{\beta} \overline{\gamma}} - b {\cal F}^{\overline{\alpha} \overline{\beta} \underline{\gamma}}\right) \Phi_{\gamma m} A_{[\mu}{}^m \partial_{\nu]}\Lambda_{\alpha \beta} \nn \\
       && - \frac 1 4 \left(a {\cal F}^{a b \overline{\alpha}} + b {\cal F}^{a b \underline{\alpha}} \right) \Phi_{\alpha m} A_{[\mu}{}^m \partial_{\nu]}\Lambda_{ab}
\ ,
       \\
\delta \widetilde M_{m n} &=& L_\xi \widetilde M_{m n} - 2 f_{p (m}{}^q \widetilde M_{n) q} \lambda^p
\ , \\
\delta \widetilde \phi &=& L_\xi\widetilde \phi + \frac 1 4\, \widetilde g^{\mu \nu}\left(\delta - L_\xi\right) \widetilde g_{\mu \nu} \ ,
\eea
where all these quantities are defined in (\ref{Ffrom})-(\ref{Fto}) and in Appendix \ref{SEC:Conventions}.

We now search for field redefinitions that eliminate the highest possible number of higher-order terms -i.e. terms that are weighted with $a$ or $b$- in the above gauge transformations. We find that defining
\bea
 \widetilde g_{\mu \nu} &=&  g_{\mu \nu} + \frac a 4 \omega^{(-)ab}_\mu \omega^{(-)}_{\mu a b} + \frac b 4 \omega^{(+)ab}_\mu \omega^{(+)}_{\mu a b}  + \frac a 4 \omega^{(-)}_{\mu}{}^{\alpha \beta} \omega^{(-)}_{\nu \alpha \beta} + \frac b 4 \omega^{(+)}_{\mu}{}^{\alpha \beta} \omega^{(+)}_{\nu \alpha \beta} \label{RedefFrom}
 \ , \\
 \widetilde A_\mu{}^m &=& A_\mu{}^m + \frac a 4 {\cal F}_{a b}{}^{\overline{\alpha}} \omega^{(-) a b}_\mu \Phi_\alpha{}^m + \frac b 4 {\cal F}_{a b}{}^{\underline{\alpha}} \omega^{(+) a b}_\mu \Phi_\alpha{}^m \nn \\
 && - \frac a 2 {\cal F}^{\alpha \beta \overline{\gamma}} \omega^{(-)}_{\mu \alpha \beta} \Phi_\gamma{}^m
 + \frac b 2 {\cal F}^{\alpha \beta \underline{\gamma}} \omega^{(+)}_{\mu \alpha \beta} \Phi_\gamma{}^m
 \ , \\
 \widetilde B_{\mu \nu} &=& B_{\mu \nu} + \widetilde A_{[\mu}{}^m A_{\nu] m}
 \ , \\
 \widetilde M_{m n} &=&  M_{m n}
 \ , \\
 \widetilde \phi &=&  \phi +  \frac 1 4 \ln \frac {\widetilde g} g\ , \label{RedefTo}
\eea
leads to the reduced transformations
\bea
\delta g_{\mu \nu} &=& L_\xi g_{\mu \nu}
\ ,\\
\delta  A_\mu{}^m &=& L_\xi  A_\mu{}^m + \partial_\mu \lambda^m + f_{p q}{}^m \lambda^p A_\mu{}^q
\ ,\\
\delta B_{\mu \nu} &=& L_\xi  B_{\mu \nu} + 2 \partial_{[\mu} \xi_{\nu]} +  A_{[\mu}{}^m \partial_{\nu]}\lambda_m - \frac 1 2 \left(a \omega^{(-)}_{[\mu}{}^{\alpha \beta} - b \omega^{(+)}_{[\mu}{}^{\alpha \beta}\right) \partial_{\nu]}\Lambda_{\alpha \beta} \nn
\ ,\\
       && - \frac 1 2 \left(a \omega_{[\mu}^{(-)ab} - b \omega_{[\mu}^{(+)ab}\right) \partial_{\nu]}\Lambda_{ab} \label{GreenSchwarz}
\ ,\\
\delta M_{m n} &=& L_\xi M_{m n} - 2 f_{p (m}{}^q  M_{n) q} \lambda^p
\ ,\\
\delta \phi &=& L_\xi \phi  \ .
\eea
We then see that the non-standard Lorentz transformations of the metric $g_{\mu \nu}$, gauge fields $A_\mu{}^m$, scalars $M_{m n}$ and dilaton $\phi$ can be totally removed.
Unsurprisingly, it turns out to be impossible to remove the dependence on $a$ and $b$ from the transformation of the two-form. If $a = b = 0$, the above would be the standard gauge transformations of all fields, where $\xi^\mu$ are vectors that parameterize the infinitesimal general coordinate transformations, $\xi_\mu$ are one-forms that generate the gauge transformations of the two-form $B_{\mu \nu}$, and $\lambda^m$ are the infinitesimal parameters of the gauge transformation of the vectors $A_\mu{}^m$ and scalars $M_{m n}$. Note that already to lowest order, we find that the gauge parameters $\lambda^m$ generate a Green-Schwarz transformation of the two-form \cite{Green:1984sg}. The additional terms in the transformation of $B_{\mu \nu}$ are also of this form, but with respect to external $H_e(\Lambda_{a b})$ and internal $H_i(\Lambda_{\alpha \beta})$ Lorentz transformations. The corresponding connections are the torsionful Lorentz spin connections $\omega^{(\pm)}_{\mu a b}$, and the internal double-Lorentz spin connections $\omega^{(\pm)}_{\mu \alpha \beta}$. These deformations induce corrections to the three-form field strength
\be
\widehat H_{\mu \nu \rho} = H_{\mu \nu \rho} - \frac 3 2 a \Omega^{(e,-)}_{\mu \nu \rho} + \frac 3 2 b \Omega^{(e,+)}_{\mu \nu \rho} - \frac 3 2 a \Omega^{(i,-)}_{\mu \nu \rho} + \frac 3 2 b \Omega^{(i,+)}_{\mu \nu \rho} \ , \label{Hhat}
\ee
in terms of the  external $\Omega^{(e,\pm)}_{\mu \nu \rho}$ and internal $\Omega^{(i,\pm)}_{\mu \nu \rho}$ Chern-Simons three-forms defined in Appendix \ref{SEC:Conventions}.

\subsection{The action}\label{sec:The action}

At this stage we  have all the information required to write an action in terms of the components of the generalized fields. Now we carry out the following procedure: we introduce the parameterizations of the generalized fluxes (\ref{Ffrom1})-(\ref{Fto2}) and flat matrices (\ref{ParamMetrics}) into the action (\ref{ActionAlphaPrime}), we expand to first order in $\alpha'$, and finally we perform the field redefinitions (\ref{RedefFrom})-(\ref{RedefTo}). The result is a complicated action whose internal and external Lorentz invariance is not manifest (but certainly a symmetry). Then some work must be done in order to bring it to a form where all the gauge symmetries are manifest. The final result is given by
\bea
S &=& \int d^n X \sqrt{-g} e^{-2 \phi} \left[ R + 4 \nabla_\mu \nabla^\mu \phi - 4 \nabla_\mu \phi \nabla^\mu \phi - \frac 1 {12} \widehat H_{\mu \nu \rho} \widehat H^{\mu \nu \rho} \right. \nn\\
&& \quad \quad \quad \quad \quad \quad \quad  \quad - \frac 1 4 F_{\mu \nu}{}^m F^{\mu \nu n} M_{m n} + \frac 1 8 \nabla_\mu M_{mn} \nabla^\mu M^{mn} - V_0 \nn \\
&&\quad \quad \quad \quad \quad \quad \quad  \quad  \left. \vphantom{\frac 1 2} + \gamma^{(-)} L^{(-)} + \gamma^{(+)} L^{(+)}\right] \ . \label{ActionGaugedSugra}
\eea

The first two lines in this expression correspond to the standard form of (gauged/half-maximal super) gravity coupled to a two-form, gauge vectors and scalar fields. The only difference is that the three-form field strength $\widehat H_{\mu \nu \rho}$ receives $\alpha'$-corrections through the external and internal Lorentz Chern-Simons terms (\ref{Hhat}). The lowest order scalar potential takes the standard form
\be
V_0 = \frac 1 {12} f_{m p}{}^r f_{n q}{}^s M^{mn} M^{pq} M_{r s} + \frac 1 4 f_{m p}{}^q f_{n q}{}^p M^{m n} + \frac 1 6 f_{mnp} f^{mnp} \ .\label{V0}
\ee
Then, the first two lines were expected and in fact could have been easily anticipated by taking into account that this is an effective action of a $G$-invariant DFT, plus the fact that the gauge transformations of the two-form are now deformed to first order in $\alpha'$ by the Green-Schwarz transformations (\ref{GreenSchwarz}).

The new piece of information here are the two terms in the last line of (\ref{ActionGaugedSugra}). For convenience we have redefined the parameters $a$ and $b$ as follows
\be
\gamma^{(\pm)} = - \frac {a \pm b} 2 \ .\label{gammapm}
\ee

These factors contain a huge number of terms and for that reason we have used Cadabra software \cite{Peeters:2007wn} in this and other computations. The explicit expressions for $L^{(\pm)}$, written in a form that makes all the gauge symmetries manifest, are

\begingroup\makeatletter\def\f@size{12}\check@mathfonts
\begin{dmath*}[compact, spread=2pt]
L^{(-)} = {\frac 1 4 \, {F}_{\mu}{}^{\nu m} {\nabla}_{\rho}{{\nabla}_{\sigma}{{F}_{\nu \epsilon m}}\, }  \left(\right. {g}^{\mu \rho}  {g}^{\sigma \epsilon} +  {g}^{\mu \sigma} {g}^{\rho \epsilon}\left.\right)
+ \frac 1 8 {\nabla}_{\mu}{{F}_{\nu \rho}{}^m}\,  {\nabla}_{\sigma}{{F}_{\gamma \epsilon m}} \left(\right.2 \, {g}^{\mu \nu} {g}^{\rho \gamma} {g}^{\sigma \epsilon} -  {g}^{\mu \sigma} {g}^{\nu \gamma} {g}^{\rho \epsilon}\left.\right)} \\ - \frac{1}{2}\, {\nabla}_{\mu}{{\widehat H}_{\nu \rho \sigma}}\,  {R}^{\mu \nu \rho \sigma} + \frac 1 4 {F}_{\mu \nu}{}^m {F}_{\rho \sigma m} {R}_{\gamma \epsilon \delta \lambda}  \left(\right. {g}^{\mu \gamma} {g}^{\nu \delta} {g}^{\rho \epsilon} {g}^{\sigma \lambda} -  {g}^{\mu \gamma} {g}^{\nu \epsilon} {g}^{\rho \delta} {g}^{\sigma \lambda} +  {g}^{\mu \rho} {g}^{\nu \gamma} {g}^{\sigma \delta} {g}^{\epsilon \lambda}\left.\right) + {F}_{\mu \nu}\,^{m} {F}^\mu{}_{\rho m} {\nabla}^{\nu}{{\nabla}^\rho{\phi}\, }  - \frac 1 4 {F}_{\mu \nu m} {\nabla}_{\rho}{{F}_{\sigma \gamma n}}\,  {\widehat H}^{\mu \gamma}{}_{\delta} {M}^{m n} \left(\right.  {g}^{\nu \sigma} {g}^{\delta \rho} - {g}^{\nu \rho} {g}^{\delta \sigma} \left.\right) - \frac 1 {16} {F}_{\mu \nu m} {\nabla}_{\rho}{{F}_{\sigma \gamma n}}\,  {\nabla}_{\epsilon}{{M}_{p q}}\,  \left(\right. 2 {M}^{n p} {\kappa}^{m q} {g}^{\mu \rho} {g}^{\nu \sigma} {g}^{\gamma \epsilon} + {M}^{n p} {\kappa}^{m q} {g}^{\mu \sigma} {g}^{\nu \gamma} {g}^{\rho \epsilon} -2 {M}^{m p} {\kappa}^{n q} {g}^{\mu \rho} {g}^{\nu \sigma} {g}^{\gamma \epsilon} - {M}^{m p} {\kappa}^{n q} {g}^{\mu \sigma} {g}^{\nu \gamma} {g}^{\rho \epsilon}\left.\right) + {F}_{\mu \nu}{}^m {\nabla}_{\rho}{{F}_{\sigma \gamma m}}\,  {\nabla}_{\epsilon}{\phi}  \left(\right.  {g}^{\mu \sigma} {g}^{\nu \epsilon} {g}^{\rho \gamma} - {g}^{\mu \rho} {g}^{\nu \sigma} {g}^{\gamma \epsilon} \left.\right) - \frac{1}{4}\, {\widehat H}^{\mu \nu \rho} {\widehat H}_{\mu}\,^{\sigma \gamma} {\nabla}_{\nu}{{\widehat H}_{\rho \sigma \gamma}}\,  + \frac 1 {16} {F}_{\mu \nu m} {F}_{\rho \sigma n} {F}_{\gamma \epsilon p} {F}_{\delta \lambda}{}^p {M}^{m n}  \left(\right. {g}^{\mu \gamma} {g}^{\nu \delta} {g}^{\rho \epsilon} {g}^{\sigma \lambda} - 2 \, {g}^{\mu \rho} {g}^{\nu \gamma} {g}^{\sigma \delta} {g}^{\epsilon \lambda}\left.\right) + \frac 1 {16} {F}_{\mu \nu}{}^m {F}_{\rho \sigma m} {\widehat H}_{\gamma \epsilon \delta} {\widehat H}_{\lambda \tau}{}^\delta  \left(\right. {g}^{\mu \gamma} {g}^{\nu \lambda} {g}^{\rho \epsilon} {g}^{\sigma \tau}  -  {g}^{\mu \gamma} {g}^{\nu \epsilon} {g}^{\rho \lambda} {g}^{\sigma \tau}  - {g}^{\mu \rho} {g}^{\nu \gamma} {g}^{\sigma \lambda} {g}^{\epsilon \tau} \left.\right) - \frac{1}{8}\, {F}^{\mu \nu m} {F}^{\rho \sigma n} {\nabla}_{\sigma}{{M}_{m n}}\,  {\widehat H}_{\mu \nu \rho} + \frac 1 {32} {F}_{\mu \nu m} {F}^\mu{}_{ \sigma n} {\nabla}_{\gamma}{{M}_{p q}}\,  {\nabla}_{\epsilon}{{M}_{r}{}^q}\,  \left(\right. {\kappa}^{m n} {\kappa}^{p r}   {g}^{\nu \gamma} {g}^{\sigma \epsilon} + 4 {\kappa}^{m p} {\kappa}^{n r}  {g}^{\nu \gamma} {g}^{\sigma \epsilon} - 2 \, {\kappa}^{m p} {\kappa}^{n r} {g}^{\nu \sigma} {g}^{\gamma \epsilon}\left.\right) - {F}^{\mu \nu m} {F}_{\mu}\,^{\rho}\,_{m} {\nabla}_{\nu}{\phi}\,  {\nabla}_{\rho}{\phi}\, +  \frac{1}{64} {\nabla}_{\mu}{{M}_{m n}}\,  {\nabla}_{\nu}{{M}_{p q}}\,  {\nabla}_{\rho}{{M}^n{}_{s}}\,  {\nabla}_{\sigma}{{M}^{q s}}\,  {M}^{m p}  \left(\right.   {g}^{\mu \rho} {g}^{\nu \sigma} -  {g}^{\mu \nu} {g}^{\rho \sigma} \left.\right) + \frac{1}{12}\, {f}_{m n p} {F}^{\mu \nu m} {F}_{\mu}\,^{\rho n} {F}_{\nu \rho}\,^{p} -  \frac{1}{8} {f}_{m}{}^{n p} {F}_{\mu \nu q} {\nabla}_{\rho}{{M}_{r n}}\,  {\nabla}_{\sigma}{{M}_{o p}}\,  {g}^{\mu \rho} {g}^{\nu \sigma} \left(\right. {M}^{q m} {\kappa}^{r o}  -  {M}^{r m} {\kappa}^{q o} \left.\right) - \frac 1 {16} {f}_{m n p} {f}_{q r}{}^p {F}_{\mu \nu o} {F}^{\mu \nu}{}_{t} \left(\right.  {M}^{o m} {M}^{t q} {M}^{n r}  + 2 {M}^{o m} {\kappa}^{t q} {\kappa}^{n r}  - {M}^{m q} {\kappa}^{o n} {\kappa}^{t r} \left.\right) - \frac 1 {32} {f}_{m n p} {f}_{q r s} {\nabla}_{\mu}{{M}_{o t}}\,  {\nabla}^\mu{{M}_{u v}} \left(\right. 2 {M}^{m q} {M}^{n r} {\kappa}^{o u} {\kappa}^{t p} {\kappa}^{v s} + {M}^{o m} {M}^{n q} {\kappa}^{t u} {\kappa}^{v r} {\kappa}^{p s} - {M}^{o m} {M}^{u n} {\kappa}^{t q} {\kappa}^{v r} {\kappa}^{p s} + 2 {\kappa}^{o m} {\kappa}^{t q} {\kappa}^{u n} {\kappa}^{v r} {\kappa}^{p s} + 2 {M}^{o m} {M}^{u q} {\kappa}^{t n} {\kappa}^{v r} {\kappa}^{p s} - {M}^{o m} {M}^{u q} {\kappa}^{t r} {\kappa}^{v n} {\kappa}^{p s} + {M}^{o u} {M}^{m q} {\kappa}^{t n} {\kappa}^{v r} {\kappa}^{p s} - 2 {\kappa}^{o u} {\kappa}^{t m} {\kappa}^{v q} {\kappa}^{n r} {\kappa}^{p s}\left.\right) - V^{(-)} \ ,
\end{dmath*}
\endgroup
where
\begingroup\makeatletter\def\f@size{12}\check@mathfonts
\begin{dmath*}[compact, spread=2pt]
V^{(-)}=  {f}_{m n p} {f}_{q r s} {f}_{o t u} {f}_{v w}{}^u {M}^{m q} \left( \right.  \frac{1}{16}\, {M}^{n r} {M}^{o v} {\kappa}^{p t} {\kappa}^{s w}  - \frac{1}{16}\,  {M}^{n o} {M}^{r v} {\kappa}^{p s} {\kappa}^{t w}  - \frac{1}{24}\, {M}^{n o} {M}^{r t} {\kappa}^{p v} {\kappa}^{s w}  - \frac{1}{8}\, {M}^{n o} {M}^{r v} {\kappa}^{p w} {\kappa}^{s t}  + \frac{1}{8}\,  {\kappa}^{n o} {\kappa}^{p t} {\kappa}^{r v} {\kappa}^{s w}  - \frac{1}{16}\,  {M}^{n o} {M}^{p v} {M}^{r t} {M}^{s w}  + \frac{7}{24}\,  {M}^{n o} {M}^{p v} {\kappa}^{r t} {\kappa}^{s w}  - \frac{1}{16}\, {\kappa}^{n o} {\kappa}^{p v} {\kappa}^{r t} {\kappa}^{s w} +\frac{1}{16}\,  {M}^{n r} {M}^{p o} {M}^{s v} {M}^{t w} - \frac{1}{8}\,  {M}^{n r} {M}^{p o} {\kappa}^{s v} {\kappa}^{t w}  - \frac{1}{16}\, {\kappa}^{n r} {\kappa}^{p o} {\kappa}^{s v} {\kappa}^{t w} \left. \right) \ .
\end{dmath*}
\endgroup

\noindent and

\begingroup\makeatletter\def\f@size{11.4}\check@mathfonts
\begin{dmath*}[compact, spread=2pt]
L^{(+)} = - \frac 1 4 {F}_{\mu \nu m} {\nabla}_{\rho}{{\nabla}_{\sigma}{{F}^\nu{}_{ \epsilon n}}\, }\,  {M}^{m n} \left(\right.  {g}^{\mu \rho}  {g}^{\sigma \epsilon} + {g}^{\mu \sigma}  {g}^{\rho \epsilon}\left.\right) + \frac{1}{4}\, {R}^{\mu \nu \rho \sigma} {R}_{\mu \nu \rho \sigma}  - \frac 1 4 {\nabla}_{\mu}{{F}_{\nu \rho m}}\,  {\nabla}_{\sigma}{{F}_{\gamma \epsilon n}}\,  {M}^{m n} \left(\right.  {g}^{\mu \nu} {g}^{\rho \gamma} {g}^{\sigma \epsilon} - \frac{3}{2}\, {g}^{\mu \sigma} {g}^{\nu \gamma} {g}^{\rho \epsilon}\left.\right) - \frac 1 8 {\nabla}_{\mu}{{\widehat H}_{\nu \rho \sigma}}\,  {\nabla}_{\gamma}{{\widehat H}_{\epsilon}{}^{\rho \sigma}}\,  \left(\right.  {g}^{\mu \epsilon} {g}^{\nu \gamma}  - {g}^{\mu \gamma} {g}^{\nu \epsilon} \left.\right) + \frac 1 {32} {\nabla}_{\mu}{{\nabla}_{\nu}{{M}_{m n}}\, }\,  {\nabla}_{\rho}{{\nabla}_{\sigma}{{M}_{p q}}\, }\,  \left(\right. {M}^{m p} {M}^{n q} {g}^{\mu \rho} {g}^{\nu \sigma} +  {M}^{m p} {M}^{n q} {g}^{\mu \sigma} {g}^{\nu \rho} -  {\kappa}^{m p} {\kappa}^{n q} {g}^{\mu \rho} {g}^{\nu \sigma} - {\kappa}^{m p} {\kappa}^{n q} {g}^{\mu \sigma} {g}^{\nu \rho}\left.\right)
- \frac 1 4 {F}_{\mu \nu m} {F}_{\rho \sigma n} {R}_{\gamma \epsilon \delta \lambda} {M}^{m n} \left(\right.  {g}^{\mu \gamma} {g}^{\nu \delta} {g}^{\rho \epsilon} {g}^{\sigma \lambda} + {g}^{\mu \gamma} {g}^{\nu \epsilon} {g}^{\rho \delta} {g}^{\sigma \lambda} + {g}^{\mu \rho} {g}^{\nu \gamma} {g}^{\sigma \delta} {g}^{\epsilon \lambda}\left.\right)
 + \frac{1}{2}\, {F}^{\mu \nu m} {F}^{\rho \sigma}\,_{m} {\nabla}_{\mu}{{\widehat H}_{\nu \rho \sigma}}\,
 + \frac 1 8 {F}_{\mu \nu m} {F}^{\mu \sigma}{}_{ n} {\nabla}^{\nu}{{\nabla}_{\sigma}{{M}_{p q}}\, }\left(\right. {M}^{m p} {M}^{n q} -  {\kappa}^{m p} {\kappa}^{n q}\left.\right) - {F}^{\mu \nu}\,_{m} {F}_{\mu}\,^{\rho}\,_{n} {\nabla}_{\nu}{{\nabla}_{\rho}{\phi}\, }\,  {M}^{m n}
 - \frac{1}{4}{F}_{\mu \nu m} {\nabla}_{\rho}{{F}_{\sigma \gamma}{}^m}\,  {\widehat H}^{\mu \gamma}{}_{\delta}  \left(\right.  {g}^{\nu \sigma} {g}^{\delta \rho} +   {g}^{\nu \rho} {g}^{\delta \sigma} \left.\right)
 + {F}_{\mu \nu m} {\nabla}_{\rho}{{F}_{\sigma \gamma n}}\,  {\nabla}_{\epsilon}{{M}^{m n}}\,   \left(\right. - \frac{3}{4}\, {g}^{\mu \rho} {g}^{\nu \sigma} {g}^{\gamma \epsilon} + \frac{1}{2}\, {g}^{\mu \sigma} {g}^{\nu \epsilon} {g}^{\rho \gamma} + \frac{3}{8}\, {g}^{\mu \sigma} {g}^{\nu \gamma} {g}^{\rho \epsilon}\left.\right) + {F}_{\mu \nu m} {\nabla}_{\rho}{{F}_{\sigma \gamma n}}\,  {\nabla}_{\epsilon}{\phi}\,  {M}^{m n} \left(\right.{g}^{\mu \rho} {g}^{\nu \sigma} {g}^{\gamma \epsilon} - {g}^{\mu \sigma} {g}^{\nu \epsilon} {g}^{\rho \gamma}\left.\right) - \frac{1}{4}\, {\widehat H}^{\mu \nu \rho} {\widehat H}_{\mu}\,^{\sigma \gamma} {R}_{\nu \sigma \rho \gamma}
 + \frac 1 {32} {F}_{\mu \nu m} {F}_{\rho \sigma n} {F}_{\gamma \epsilon p} {F}_{\delta \lambda q} \left(\right. {M}^{m n} {M}^{p q} {g}^{\mu \gamma} {g}^{\nu \delta} {g}^{\rho \epsilon} {g}^{\sigma \lambda} + 2\, {M}^{m n} {M}^{p q} {g}^{\mu \gamma} {g}^{\nu \epsilon} {g}^{\rho \delta} {g}^{\sigma \lambda} + 8\, {M}^{m n} {M}^{p q} {g}^{\mu \rho} {g}^{\nu \gamma} {g}^{\sigma \delta} {g}^{\epsilon \lambda} - 7\, {\kappa}^{m n} {\kappa}^{p q} {g}^{\mu \gamma} {g}^{\nu \delta} {g}^{\rho \epsilon} {g}^{\sigma \lambda} + 2\, {\kappa}^{m n} {\kappa}^{p q} {g}^{\mu \gamma} {g}^{\nu \epsilon} {g}^{\rho \delta} {g}^{\sigma \lambda}\left.\right)
+ \frac 1 {16}{F}_{\mu \nu m} {F}_{\rho \sigma n} {\widehat H}_{\gamma \epsilon \delta} {\widehat H}_{\lambda \tau}{}^\delta {M}^{m n} \left(\right. {g}^{\mu \gamma} {g}^{\nu \epsilon} {g}^{\rho \lambda} {g}^{\sigma \tau}  +  {g}^{\mu \gamma} {g}^{\nu \lambda} {g}^{\rho \epsilon} {g}^{\sigma \tau} + 3\, {g}^{\mu \rho} {g}^{\nu \gamma} {g}^{\sigma \lambda} {g}^{\epsilon \tau} \left.\right)
- \frac 1 {16} {F}_{\mu \nu m} {F}_{\rho \sigma}{}^n {\nabla}_{\gamma}{{M}_{p n}}\,  {\widehat H}_{\epsilon \delta \lambda} {M}^{m p}  \left(\right. 3\, {g}^{\mu \epsilon} {g}^{\nu \delta} {g}^{\rho \lambda} {g}^{\sigma \gamma} - 3\, {g}^{\mu \epsilon} {g}^{\nu \gamma} {g}^{\rho \delta} {g}^{\sigma \lambda} - 2 \, {g}^{\mu \rho} {g}^{\nu \epsilon} {g}^{\sigma \delta} {g}^{\lambda \gamma}\left.\right)
- \frac{3}{32} {F}_{\mu \nu m} {F}_{\rho \sigma n} {\nabla}_{\gamma}{{M}_{p q}}\,  {\nabla}_{\epsilon}{{M}_{r s}}\,  \left(\right.  {M}^{m n} {\kappa}^{p r} {\kappa}^{q s} {g}^{\mu \rho} {g}^{\nu \gamma} {g}^{\sigma \epsilon} + 3\, {M}^{m p} {\kappa}^{n r} {\kappa}^{q s} {g}^{\mu \rho} {g}^{\nu \gamma} {g}^{\sigma \epsilon} + 2 \, {M}^{m p} {\kappa}^{n r} {\kappa}^{q s} {g}^{\mu \rho} {g}^{\nu \sigma} {g}^{\gamma \epsilon} - {M}^{p r} {\kappa}^{m q} {\kappa}^{n s} {g}^{\mu \rho} {g}^{\nu \gamma} {g}^{\sigma \epsilon} - \frac 2 3 \, {M}^{p r} {\kappa}^{m q} {\kappa}^{n s} {g}^{\mu \rho} {g}^{\nu \sigma} {g}^{\gamma \epsilon}\left.\right)
- {F}^{\mu \nu m} {F}_{\mu}\,^{\rho n} {\nabla}_{\nu}{{M}_{m n}}\,  {\nabla}_{\rho}{\phi}\,  + {F}^{\mu \nu}\,_{m} {F}_{\mu}\,^{\rho}\,_{n} {\nabla}_{\nu}{\phi}\,  {\nabla}_{\rho}{\phi}\,  {M}^{m n}
- \frac{1}{32} {\widehat H}_{\mu \nu \rho} {\widehat H}^\mu{}_{\gamma \epsilon} {\widehat H}_{\delta \lambda \tau} {\widehat H}_{\xi \zeta}{}^\tau \left(\right.  {g}^{\nu \delta} {g}^{\rho \xi} {g}^{\gamma \lambda} {g}^{\epsilon \zeta}  - {g}^{\nu \gamma} {g}^{\rho \delta} {g}^{\epsilon \xi} {g}^{\lambda \zeta} \left.\right)
- \frac{1}{32}\, {\nabla}_{\mu}{{M}^{m n}}\,  {\nabla}_{\nu}{{M}_{m n}}\,  {\widehat H}^{\rho \sigma \mu} {\widehat H}_{\rho \sigma}\,^{\nu} + {\nabla}_{\mu}{{M}_{m n}}\,  {\nabla}_{\nu}{{M}^m{}_{q}}\,  {\nabla}_{\rho}{{M}_{r s}}\,  {\nabla}_{\sigma}{{M}_{o}{}^s}\,  \left(\right.\frac{1}{128}\,  {\kappa}^{n q} {\kappa}^{r o} {g}^{\mu \rho} {g}^{\nu \sigma} + \frac{1}{16}\, {\kappa}^{n r} {\kappa}^{q o}  {g}^{\mu \nu} {g}^{\rho \sigma} - \frac{3}{64}\,  {\kappa}^{n r} {\kappa}^{q o} {g}^{\mu \sigma} {g}^{\nu \rho}\left.\right)
- \frac{1}{24} {f}_{m n p} {F}_{\mu \nu q} {F}^{\mu \sigma}{}_{r} {F}^\nu{}_{\sigma s}  {M}^{q m} \left(\right.  {M}^{r n} {M}^{s p} + 9 \,  {\kappa}^{r n} {\kappa}^{s p}\left.\right)
- \frac{1}{8}\, {f}_{m n}{}^p {F}_{\mu \nu q} {\nabla}_{\rho}{{M}_{r p}}\,  {\widehat H}^{\mu \nu \rho} {M}^{q m} {M}^{r n}
\end{dmath*}
\begin{dmath*}[compact, spread=2pt]
+ \frac 1 {32} {f}_{m n}{}^p {F}_{\mu \nu q} {\nabla}_{\rho}{{M}_{r s}}\,  {\nabla}_{\sigma}{{M}_{o p}}\,  {g}^{\mu \rho} {g}^{\nu \sigma} \left(\right.3 {M}^{q m} {M}^{r n} {\kappa}^{s o}  -  {M}^{q m} {M}^{r o} {\kappa}^{s n}  + 16\, {\kappa}^{q m} {\kappa}^{r o} {\kappa}^{s n}  - 12\, {M}^{r m} {M}^{o n} {\kappa}^{q s} \left.\right)
- \frac 1 {16} {f}_{m n p} {f}_{q r s} {F}_{\mu \nu o} {F}^{\mu \nu}{}_{t}  \left(\right. 2\, {M}^{o m} {M}^{n q} {\kappa}^{t r} {\kappa}^{p s} -2\, {M}^{o m} {M}^{t q} {M}^{n r} {M}^{p s} + {M}^{o m} {M}^{t q} {\kappa}^{n r} {\kappa}^{p s} + {\kappa}^{o m} {\kappa}^{t q} {\kappa}^{n r} {\kappa}^{p s}\left.\right)
+ \frac 1 {32} {f}_{m n p} {f}_{q r s} {\nabla}_{\mu}{{M}_{o t}}\,  {\nabla}^{\mu}{{M}_{u v}}\,   \left(\right.14\, {M}^{m q} {\kappa}^{o n} {\kappa}^{t r} {\kappa}^{u p} {\kappa}^{v s} + 3 \, {M}^{o m} {\kappa}^{t u} {\kappa}^{v q} {\kappa}^{n r} {\kappa}^{p s} - 6\, {M}^{o m} {M}^{n q} {M}^{p r} {\kappa}^{t u} {\kappa}^{v s} +  {M}^{o m} {M}^{u n} {M}^{p q} {\kappa}^{t r} {\kappa}^{v s} + 2\, {M}^{o m} {M}^{u q} {M}^{n r} {\kappa}^{t p} {\kappa}^{v s} + {M}^{o m} {M}^{u q} {M}^{n r} {\kappa}^{t s} {\kappa}^{v p} + 2\, {M}^{o u} {M}^{m q} {M}^{n r} {\kappa}^{t p} {\kappa}^{v s} -  {M}^{o u} {\kappa}^{t m} {\kappa}^{v q} {\kappa}^{n r} {\kappa}^{p s}\left.\right) - V^{(+)} \ ,
\end{dmath*}
\endgroup
\noindent where
\begingroup\makeatletter\def\f@size{12}\check@mathfonts
\begin{dmath*}[compact, spread=2pt]
V^{(+)}= {f}_{m n p} {f}_{q r s} {f}_{o t u} {f}_{v x w} \left(\right.\frac{1}{96}\, {M}^{m q} {M}^{n o} {M}^{p v} {M}^{r t} {M}^{s w} {M}^{u x} + \frac{1}{8}\, {M}^{m q} {M}^{n o} {M}^{p v} {M}^{r t} {\kappa}^{s w} {\kappa}^{u x} + \frac{1}{8}\, {M}^{m q} {M}^{n o} {\kappa}^{p v} {\kappa}^{r t} {\kappa}^{s w} {\kappa}^{u x} - \frac{7}{96}\, {\kappa}^{m q} {\kappa}^{n o} {\kappa}^{p v} {\kappa}^{r t} {\kappa}^{s w} {\kappa}^{u x} + \frac{1}{16}\, {M}^{m q} {M}^{n r} {M}^{p o} {M}^{s v} {M}^{t w} {M}^{u x} - \frac{1}{16}\, {M}^{m q} {M}^{n r} {M}^{p o} {M}^{s v} {\kappa}^{t w} {\kappa}^{u x} - \frac{1}{16}\, {M}^{m q} {M}^{n r} {\kappa}^{p o} {\kappa}^{s v} {\kappa}^{t w} {\kappa}^{u x} + \frac{1}{8}\, {M}^{m q} {M}^{n r} {M}^{p o} {M}^{t v} {\kappa}^{s w} {\kappa}^{u x} - \frac{1}{8}\, {M}^{m q} {M}^{n o} {\kappa}^{p r} {\kappa}^{s v} {\kappa}^{t w} {\kappa}^{u x} + \frac{1}{16}\, {M}^{m q} {M}^{n r} {M}^{o v} {M}^{t w} {\kappa}^{p u} {\kappa}^{s x} - \frac{7}{32}\, {M}^{m q} {M}^{n o} {M}^{r v} {M}^{t w} {\kappa}^{p x} {\kappa}^{s u} + \frac{1}{32}\, {M}^{m q} {M}^{o v} {\kappa}^{n t} {\kappa}^{p w} {\kappa}^{r u} {\kappa}^{s x}\left.\right) \ .
\end{dmath*}
\endgroup

We have explicitly written here only terms that are first order in $\alpha'$, but higher order terms   are included in the deformation of the three-form field strength $\widehat H_{\mu \nu \rho} = H_{\mu \nu \rho} + {\cal O}(\alpha')$ so as to ensure Lorentz invariance of the action. This form of the action can be simplified through Bianchi identities like those discussed in Appendix \ref{SEC:Conventions}, field redefinitions as discussed in Appendix \ref{SEC:Redefs}, and integrations by parts. We postpone this task to Subsection \ref{SEC:Simplifications} and now move on to study some simple special cases of relevance, such as the bosonic and the heterotic string low energy effective actions.

\subsection{The bosonic string}

Let us now take the specifications required to make contact with the bosonic string. In this case,  $n = 26$, $d = 0$ and $N = 0$ and the values of the parameters are $(a , b) = (- \alpha' , - \alpha')$. Since we are truncating the internal part of the action, we can simply set $A_\mu{}^m = 0$, $f_{m n p} = 0$ and consider a trivial scalar frame $\Phi_m{}^\alpha = \delta_m{}^\alpha$. Evaluating the action (\ref{ActionGaugedSugra}) in this form, one rapidly arrives at
\bea
S_{Bos} &=& \int d^{26} X \sqrt{- g} e^{- 2 \phi} \left[ R + 4 \nabla_\mu \nabla^\mu \phi - 4 \nabla_\mu \phi \nabla^\mu \phi - \frac 1 {12} \widehat H_{\mu \nu \rho} \widehat H^{\mu \nu \rho} \right. \nn \\
&& \quad \quad  \quad  \quad \quad  \quad  \quad \quad  \quad \left. - \frac {\alpha'} 8 R^{(-)}_{\mu \nu a}{}^b R^{(-)\mu \nu}{}_b{}^a - \frac {\alpha'} 8 R^{(+)}_{\mu \nu a}{}^b R^{(+)\mu \nu}{}_b{}^a \right] \ ,
\eea
where
\be
\widehat H_{\mu \nu \rho} = 3 \partial_{[\mu} B_{\nu \rho]} + \frac 3 2 \alpha' \Omega_{\mu \nu \rho}^{(e,-)} - \frac 3 2 \alpha' \Omega_{\mu \nu \rho}^{(e,+)} \ .
\ee
Written in this form, it exactly coincides with the form of the bosonic string as displayed in \cite{Marques:2015vua}. There, it was shown that decomposing the Riemann tensor and Chern-Simons terms by separating the torsion part of the spin connection, and performing some field redefinitions, this action matches the standard bosonic string effective action obtained by Metsaev-Tseytlin \cite{Metsaev:1987zx}
\bea
S_{Bos} &=& \int d^{26} X \sqrt{- g} e^{- 2 \phi} \left[ R + 4 \nabla_\mu \nabla^\mu \phi - 4 \nabla_\mu \phi \nabla^\mu \phi - \frac 1 {12}  H_{\mu \nu \rho} H^{\mu \nu \rho} \right.  \\
&& \quad \quad  \quad  \quad \quad  \quad  \quad \left. + \frac {\alpha'} 4 \left( R_{\mu \nu \rho \sigma} R^{\mu \nu \rho \sigma} - \frac 1 2 H^{\mu \nu \rho} H_{\mu \sigma \lambda} R_{\nu \rho}{}^{\sigma \lambda} + \frac 1 {24} H^4 - \frac 1 8 H^2_{\mu \nu} H^{2 \mu \nu}\right)\right] \ , \nn
\eea
where
\bea
 H_{\mu \nu \rho} &=& 3 \partial_{[\mu} B_{\nu \rho]}
 \ ,\\
 H^2_{\mu \nu} &=& H_{\mu \rho \sigma} H_\nu{}^{\rho \sigma}
 \ ,\\
H^4 &=& H^{\mu \nu \rho} H_{\mu \sigma}{}^\lambda H_{\nu \lambda}{}^\delta H_{\rho \delta}{}^\sigma \ .
\eea

\subsection{The heterotic string}
We now move on to the heterotic string. Since we are including gauge vectors with a non-Abelian gauge group, the action (\ref{ActionGaugedSugra}) is expected to give rise to extra terms with respect to the results in \cite{Marques:2015vua}. We will show that such extra terms are exactly those required to match the Bergshoeff-de Roo action \cite{Bergshoeff:1988nn}. First we have to take $n= 10$, $d = 0$ and $N = 496$. Then, considering that the gauge group induced by the gaugings is either  $SO(32)$ or $E_8\times E_8$, the $f_{m n}{}^p$ must be taken to match the structure constants of these groups. In addition, since there are no scalar fields in the heterotic string apart from the dilaton, we have to trivialize the scalar frame $\Phi_m{}^\alpha = \delta_m{}^\alpha$. We realize that  we have to set $M_{m n} = \eta_{m n}$, where $\eta_{m n}$ is the Killing metric of the gauge group, so that the zeroth order part of the action (\ref{ActionGaugedSugra})  matches that in \cite{Bergshoeff:1988nn}. In addition, since the gaugings are now the structure constants of the gauge group, we have
\be
f_{m p}{}^q f_{n q}{}^p = \gamma \eta_{m n} \ ,
\ee
for some constant $\gamma$. The parameters must be set to $(a,b) = (- \alpha', 0)$ as in \cite{Marques:2015vua}.

To lowest order, the two derivative action contains even powers of $\kappa_{m n}$, and then there are  two options to relate it to the Killing metric $\eta_{m n}$,
\be
(H_+)\ \ \ \kappa_{m n} = \eta_{m n} \ , \ \ \ \ \ (H_-)\ \ \ \kappa_{m n} = - \eta_{m n} \ . \label{Hpm}
\ee
We then explore the cases $H_+$ and $H_-$ to first order in $\alpha'$ separately.

Let us begin with $H_+$. Taking a close look into $\widehat H_{\mu \nu \rho}$, the dependence on $\Omega^{(i,-)}_{\mu \nu \rho}$ in (\ref{widehatH}) trivially cancels because the choices we have made set $P_{\alpha \beta} = 0$ in (\ref{internalPs1}). Then, in this case the three-form field strength (\ref{Hhat}) takes the form
\be
\widehat H_{\mu \nu \rho} = H_{\mu \nu \rho} + \frac 3 2 \alpha' \Omega^{(e,-)}_{\mu \nu \rho}  \ .
\ee
It is easy to see that performing some field redefinitions, integrations by parts, and using Bianchi identities, the action can be taken to the form
\bea
S_{H_+} &=& \int d^{10} X \sqrt{- g} e^{- 2 \phi} \left[ R + 4 \nabla_\mu \nabla^\mu \phi - 4 \nabla_\mu \phi \nabla^\mu \phi - \frac 1 {12} \widehat H_{\mu \nu \rho} \widehat H^{\mu \nu \rho} - \frac 1 4 F_{\mu \nu m} F^{\mu \nu m}\right. \nn \\
&& \quad \quad  \quad  \quad  + \frac {\alpha'} 8 \left(\, {R}^{(-)\mu \nu \rho \sigma} {R}^{(-)}_{\mu \nu \rho \sigma} - \, {R}^{(-)}_{\mu \nu \rho \sigma} {F}^{\mu \nu m} {F}^{\rho \sigma}\,_{m} + \frac{1}{2}\, {F}^{\mu \nu m} {F}^{\rho \sigma}\,_{m} {F}_{\mu \rho}\,^{n} {F}_{\nu \sigma n}\right. \nn \\
&&\quad \quad  \quad  \quad \quad \quad \quad \left. \left. - \frac{1}{2}\, {F}^{\mu \nu m} {F}_{\mu}\,^{\rho}\,_{m} {F}_{\nu}\,^{\sigma n} {F}_{\rho \sigma n} + \frac{2}{3}\, {F}^{\mu \nu m} {F}_{\mu}\,^{\rho n} {F}_{\nu \rho}\,^{p} {f}_{m n p} \right) \right] \ .
\eea

Let us now move on to the case $H_-$. First we note that in this case the dependence of $\widehat H_{\mu \nu \rho}$ on $\Omega^{(i,-)}_{\mu \nu \rho}$ in (\ref{widehatH}) does not cancel because the choices we have made set $P_{\alpha \beta} = - \eta_{\alpha \beta}$ in (\ref{internalPs1}). Then, it can be checked that due to the choice $\Phi_m{}^\alpha = \delta_m{}^\alpha$, the internal Lorentz spin connection $\omega^{(-)}_{\mu \alpha}{}^{\beta}$ is given by\footnote{Note that both sides of the equality seem to transform differently. The anomalous part of the transformation of the LHS is  $\delta \omega^{(-)}_{\mu \alpha}{}^\beta = \partial_\mu \Lambda_\alpha{}^\beta$ while the RHS transforms as $\delta \left(f_{m \alpha}{}^\beta A_{\mu}{}^m \right) = f_{m \alpha}{}^\beta \partial_\mu \lambda^m$ up to a covariant contribution. This apparent inconsistency is resolved by noting that, because we have fixed $\Phi_m{}^\alpha$ to a constant, we also need to gauge fix the internal Lorentz symmetry by identifying $\Lambda_{\alpha \beta} = f_{m \alpha \beta} \lambda^m$.}
\be
\omega^{(-)}_{\mu \alpha}{}^\beta = A_\mu{}^m f_{m \alpha}{}^\beta  \ , \label{identification}
\ee
and then
\be
\Omega^{(i,-)}_{\mu \nu \rho} = \gamma \Omega^{(g)}_{\mu \nu \rho} \ .
\ee
We then see that the three-form curvature becomes (note that changing the identification $\kappa_{m n} = - \eta_{m n}$ changes the sign of $\Omega^{(g)}_{\mu \nu \rho}$)
\be
\widehat H_{\mu \nu \rho} = 3 \partial_{[\mu}{ B_{\nu \rho]}} + 3 \beta \, \Omega^{(g)}_{\mu \nu \rho} + \frac 3 2 \alpha' \Omega^{(e,-)}_{\mu \nu \rho}  \ .
\ee
with $\beta = 1 + \frac 1 2 \alpha' \gamma$. We can now compute the action for the choice $H_-$ by specifying (\ref{ActionGaugedSugra}) for this particular case. After some field redefinitions, integrations by parts and using Bianchi identities, we find
\bea
S_{H_-} &=& \int d^{10} X \sqrt{- g} e^{- 2 \phi} \left[ R + 4 \nabla_\mu \nabla^\mu \phi - 4 \nabla_\mu \phi \nabla^\mu \phi - \frac 1 {12} \widehat H_{\mu \nu \rho} \widehat H^{\mu \nu \rho} - \frac \beta 4 F_{\mu \nu m} F^{\mu \nu m}\right. \nn \\
&& \quad \quad  \quad  \quad \quad \quad \quad \quad  \left.+ \frac {\alpha'} 8 \left(\, {R}^{(-)\mu \nu \rho \sigma} {R}^{(-)}_{\mu \nu \rho \sigma}  - \frac 1 2 T_{\mu \nu} T^{\mu \nu} - \frac 3 2 T_{\mu \nu \rho \sigma} T^{\mu \nu \rho \sigma}\right) \right] \ ,
\eea
where, following \cite{Bergshoeff:1988nn} we have defined
\be
T_{\mu \nu} = F_{\mu}{}^{\rho m} F_{\rho \nu m} \ , \ \ \ \ \ T_{\mu \nu \rho \sigma} = F_{[\mu \nu}{}^m F_{\rho \sigma] m} \ .
\ee
The $\beta$-dependence can be eliminated through a shift in the gauge fields and gaugings. Written in this form, it can be checked that some further rescalings can be performed in order to match the heterotic action by Bergshoeff and de Roo \cite{Bergshoeff:1988nn} exactly.

We then conclude that the heterotic string effective action is obtained from the choice $H_-$. The effective action that results from the choice $H_+$, although not related to string theory, still enjoys an underlying duality structure. The difference between both theories are the Buscher rules \cite{Buscher:1987qj} with respect to which they are invariant. The duality covariant fields (that we have denoted with tildes) are related in a different way to the gauge covariant fields on which these actions depend. In fact, it can be seen from (\ref{RedefFrom})-(\ref{RedefTo}) that the relations between duality and gauge covariant fields depend on $\kappa_{\alpha \beta}$, and hence on the choice (\ref{Hpm}).

\subsection{Higher-derivative half-maximal gauged supergravity} \label{SEC:Simplifications}

In Section \ref{sec:The action} we gave the explicit expression of the gauged $\alpha'$-deformed DFT action and we showed that the first order $\alpha'$-corrections are contained in ${\small - \frac 1 {12}}\widehat H_{\mu\nu\rho}\widehat H^{\mu\nu\rho}$ and in the last line of (\ref{ActionGaugedSugra}). This action can be further simplified performing several manipulations, which include Bianchi identities, field redefinitions and integrations by parts. It would be desirable to take the action to a minimal form. In this section we display some partial simplifications, and the interested reader can find the technical details in Appendix \ref{AppSF}. Let us note that although the title of this section refers to gauged supergravities, the results are more general and apply to arbitrary values of the parameters $a$ and $b$. Since the case $b = 0$ captures the first-order heterotic string corrections, we believe that this choice corresponds to the corrections that admit a supersymmetric completion. One must then keep in mind that the corrections to half-maximal gauged supergravities correspond to the choice $b = 0$, although we will be general and discuss the generic case.

In order to have a more compact form of the action, it is  useful to reorganize it in terms of the parameters $a$ and $b$ instead of $\gamma^{(\pm)}$, i.e.
we introduce the calligraphic ${\mathcal L}^{(\pm)}$ as follows
\bea
S &=& \int d^n X \sqrt{-g} e^{-2 \phi} \left[ R + 4 \nabla_\mu \nabla^\mu \phi - 4 \nabla_\mu \phi \nabla^\mu \phi - \frac 1 {12} \widehat H_{\mu \nu \rho} \widehat H^{\mu \nu \rho} \right. \nn\\
&& \quad \quad \quad \quad \quad \quad \quad  \quad - \frac 1 4 F_{\mu \nu}{}^m F^{\mu \nu n} M_{m n} + \frac 1 8 \nabla_\mu M_{mn} \nabla^\mu M^{mn} - V_0 \nn \\
&&\quad \quad \quad \quad \quad \quad \quad  \quad  \left. \vphantom{\frac 1 2} +a\, {\cal L}^{(-)}+ b\,{\cal L}^{(+)}\right] \ ,
\eea
This is just a rewriting of (\ref{ActionGaugedSugra}) with the identifications
\begin{eqnarray}
\gamma^{(-)}L^{(-)}+\gamma^{(+)}L^{(+)} \;=\;a\, {\cal L}^{(-)}+ b\,{\cal L}^{(+)}\ . \label{leff}
\end{eqnarray}
It can be shown that these corrections take the form
\begin{eqnarray}
{\mathcal L}^{(\pm)}=  \frac{1}{8}\, {\tilde{R}{}^{(\pm)}}^{\mu \nu \rho \sigma} {\tilde{R}{}^{(\pm)}}_{\mu \nu \rho \sigma}
+ {\mathcal L}_{ungauged}^{(\pm)} + {\mathcal L}_{gauged}^{(\pm)} - {\cal V}^{(\pm)}\ ,\label{LeffSF}
\end{eqnarray}
where $\tilde R^{(\pm)}_{\mu \nu \rho \sigma}$ is defined as follows
\begin{eqnarray}
\tilde{R}^{(\pm)}_{\mu \nu \rho \sigma}= R^{(\pm)}_{\mu \nu \rho \sigma} \pm \frac12 F_{\mu \nu m} F_{\rho \sigma n}
(P^{(\pm)}{}^{m n}-2\, P^{(\mp)}{}^{m n}) \ ,
\end{eqnarray}
 and the corrections to the scalar potential $\gamma^{(-)}V^{(-)}+\gamma^{(+)}V^{(+)} \;=\;a\, {\cal V}^{(-)}+ b\,{\cal V}^{(+)}$
are explicitly given by
\begin{eqnarray}
{\cal V}^{(\pm)}&=& \left( P^{(\pm)}_{m m'} P^{(\pm)}_{n n'} P^{(\mp)}_{p p'}
					- P^{(\mp)}_{m m'} P^{(\mp)}_{n n'} P^{(\pm)}_{p p'} \right)
					P^{(\pm)}_{q q'} P^{(\pm)}_{r r'} P^{(\mp)}_{s s'}
					 f^{m p q} f^{n p' q'} f^{m' r s} f^{n'r's'} \label{calV} \nn\\
					&+& \left(  P^{(\pm)}_{m m'} P^{(\pm)}_{n n'} P^{(\mp)}_{p p'}
					             + \frac43 P^{(\mp)}_{m m'} P^{(\mp)}_{n n'} P^{(\pm)}_{p p'} \right)
					P^{(\pm)}_{q q'} P^{(\pm)}_{r r'} P^{(\mp)}_{s s'}
					  f^{m n s} f^{m' p r} f^{n' p' q} f^{q' r' s'}\ , \ \ \ \ \ \ \label{efpot}
\end{eqnarray}
where we have conveniently renamed  the projectors $P = P^{(-)}$ and $\bar P = P^{(+)}$.
The terms in ${\mathcal L}_{gauged}^{(\pm)}$ include the higher-derivative interactions that explicitly depend on the gaugings $f_{m n p}$, while ${\mathcal L}_{ungauged}^{(\pm)}$ contains the terms that only depend implicitly on the gaugings through the field strengths and the covariant derivatives. Their explicit expressions are given by
\begin{eqnarray}
{\mathcal L}_{ungauged}^{(\pm)}&=&
- \frac{1}{256}\, {\nabla}_{\mu}{{M}_{m n}}\,  {\nabla}_{\nu}{{M}_{p q}}\,  {\nabla}^{\mu}{{M}^{p q}}\,  {\nabla}^{\nu}{{M}^{m n}}\,
- \frac{1}{128}\, {\nabla}_{\mu}{{M}_{m n}}\,  {\nabla}_{\nu}{{M}^{m p}}\,  {\nabla}^{\mu}{{M}_{p q}}\,  {\nabla}^{\nu}{{M}^{n q}}\,  \cr
&-& \frac{1}{64}\, {\nabla}_{\mu}{{M}_{n p}}\,  {\nabla}_{\nu}{{M}_{q r}}\,  {\nabla}^{\mu}{{M}^{p r}}\,  {\nabla}^{\nu}{{M}^{m q}}\,
 (4\, {\kappa}_{m}\,^{n} \mp {M}_{m}\,^{n})
+ \frac{1}{16}\, {\nabla}_{(\mu}{{\nabla}_{\nu)}{{M}_{m n}}\, }\,  {\nabla}^{\mu}{{\nabla}^{\nu}{{M}^{m n}}\, }\,   \cr
&+& \frac{1}{128} {F}_{\mu \nu m} {F}_{\rho \sigma n} {F}^{\mu \nu}\,_{p} {F}^{\rho \sigma}\,_{q}
\left({\kappa}^{m n} {\kappa}^{p q}-13\,M^{m n} M^{p q}\pm 12\,\kappa^{m n} M^{p q}   \right) \cr
&-&\frac{1}{64} {F}_{\mu \nu m} {F}_{\rho \sigma n} {F}^{\mu \rho}\,_{p} {F}^{\nu \sigma}\,_{q}
( M^{m n} M^{p q}\pm 4\, M^{m n} \kappa^{p q} - \kappa^{m n} \kappa^{p q} + 4 \, {M}^{m p} {M}^{n q} ) \cr
&\pm& \frac{1}{4}{\nabla}^{\rho}{\nabla}_{\rho}\left({P^{(\pm)}}^{m n} {F}_{\mu \nu m} {F}^{\mu \nu}\,_{n}\right)
- \frac{1}{8}\, M^{m n} {\nabla}_{\rho}{{F}_{\mu \nu m}}\,  {\nabla}^{\rho}{{F}^{\mu \nu}\,_{n}}\cr
&\pm& \frac{1}{4} {P^{(\pm)}}^{m n} {\nabla}^{\mu}{\nabla}_{\rho}\left({F}_{\mu \nu m} {F}^{\nu \rho}\,_{n}\right)
+\frac{1}{16}\, {F}_{\mu \nu m} {F}^{\nu \rho}\,_{n} {\nabla}^{\mu}{{\nabla}_{\rho}{{M}_{p q}}\, }\,
\left(M^{m p} M^{n q}-\kappa^{m p} \kappa^{n q}\right) \cr
&-& \frac{1}{32}\, {F}_{\mu \nu m} {F}^{\nu \rho}\,_{n} {\nabla}^{\mu}{{M}^{p q}}\, {\nabla}_{\rho}{{M}_{p q}}\, {M}^{m n}
 \mp \frac{1}{32}{F}_{\mu \nu m} {F}^{\mu \nu}\,_{n} {\nabla}^{\rho}{{M}^{m p}}\,  {\nabla}_{\rho}{{M}^{n q}}\,
( {\kappa}_{p q} \pm 4\, {M}_{p q}) \cr
&+&\frac{1}{8}\, {F}_{\mu \nu m} {F}^{\nu \rho}\,_{n} {\nabla}^{\mu}{{M}^{m p}}\, {\nabla}_{\rho}{{M}^{n q}}\,
(3\,M_{p q} \mp \kappa_{p q}) \cr
&-& \frac{3}{8}\,{F}_{\mu \nu m} {\nabla}_{\rho}{{F}^{\mu \nu}\,_{n}}\,  {\nabla}^{\rho}{{M}^{m n}}+ \frac{1}{4}\, {F}_{\mu \nu m} {\nabla}^{\mu}{{M}^{m n}}\,  {\nabla}_{\rho}{{F}^{\nu \rho}\,_{n}}\,  \cr
&+& \frac{1}{64}\, {H}^{\nu \rho \sigma} {H}_{\gamma \rho \sigma}\left( {\nabla}_{\nu}{{M}_{m n}}\,  {\nabla}^{\gamma}{{M}^{m n}}\,
-4\, {F}_{\mu \nu m} {F}^{\mu \gamma}\,_{n} {M}^{m n}\right) \cr
&+& \frac{1}{16}\, {F}_{\mu \nu m} {F}_{\rho \sigma n}
\left( {H}^{\mu \rho \gamma} {H}^{\nu \sigma}\,_{\gamma} {M}^{m n}
\mp {H}^{\mu \nu \gamma} {H}^{\rho \sigma}\,_{\gamma} {P^{(\pm)}}^{m n}\right) \cr
&-& \frac{1}{16}\, {\nabla}^{\sigma}{{M}^{n}\,_{p}}\; {F}_{\mu \nu m}
\left({F}^{\mu \rho}\,_{n} {H}^{\nu}\,_{\rho \sigma} {M}^{m p}
- \, {F}_{\rho \sigma n} {H}^{\mu \nu \rho} \left(3 M^{m p} \mp \kappa^{m p} \right)  \right)\cr
&-& \frac{1}{8}\, {F}_{\mu \nu m} {H}^{\mu \rho \sigma} {\nabla}^{\nu}{{F}_{\rho \sigma n}} \,
\left(3\,\kappa^{m n} \pm M^{m n}\right)\,,
\end{eqnarray}
and
\begin{eqnarray}
{\mathcal L}_{gauged}^{(\pm)}&=&
\pm \frac{1}{8} {f}_{m n p} {f}_{q r s} {\nabla}_{\mu}{{M}^{{l} p}}\,  {\nabla}^{\mu}{{M}^{{k} s}}\,  \left[
- {P^{(\mp)}}^{m q} ({\kappa}^{r}\,_{{l}} {\kappa}^{n}\,_{{k}} + 2\, {P^{(+)}}^{r}\,_{{l}} {P^{(-)}}^{n}\,_{{k}})\right. \cr
&& \ \ \ \ \ \ \ \left.+{P^{(\pm)}}^{m q}
\left( 2\, {P^{(+)}}^{n}\,_{{l}} {P^{(-)}}^{r}\,_{{k}} + 2\, {\kappa}^{r}\,_{{l}} {\kappa}^{n}\,_{{k}}
 -{M}^{r n} {M}_{{l} {k}}
+{P^{(\mp)}}^{r n} ( {\kappa}_{{k} {l}} \pm 2\, {M}_{{k} {l}}) \right)\right] \nn\\
&&+ \frac{1}{16}\, {f}_{{k} p r} {f}_{n q s} {F}_{\mu \nu m} {F}^{\mu \nu}\,_{{l}}  {P^{(\pm)}}^{r s}
\left(\pm 2\, {M}^{m {k}} ({\kappa}^{n {l}} {\kappa}^{p q} +{M}^{n {m_1}} {M}^{p q}) \right. \cr
&&\ \ \ \ \ \ \ \ - \left.{\kappa}^{p q} ({\kappa}^{n {l}} {\kappa}^{m {k}} +{M}^{n {l}} {M}^{m {k}}) \right) \cr
&&- \frac{1}{48}\, {F}^{\mu}\,_{\gamma m} {F}_{\mu \nu n} {F}^{\gamma \nu}\,_{p} {f}_{q r s}
\left( \, {M}^{m q} ({\kappa}^{n r} {\kappa}^{p s} + {M}^{n r} {M}^{p s})
+ 2(4\, {M}^{m q} \pm {\kappa}^{m q}) {\kappa}^{n r} {\kappa}^{p s}\right) \cr
&&+ \frac{1}{4}\, {F}_{\mu \nu m} {\nabla}^{\mu}{{M}_{p r}}\,  {\nabla}^{\nu}{{M}_{q s}}\,  {f}_{n}\,^{r s}
\left(\pm  {M}^{n [m} {P^{(\mp)}}^{p] q} + {\kappa}^{m n} {\kappa}^{p q} - {M}^{m p} {M}^{n q}\right)\cr
&&- \frac{1}{16}\, {F}_{\mu \nu m} {H}^{\mu \nu \rho} {M}^{m n} {M}^{p q} {\nabla}_{\rho}{{M}_{p}\,^{s}}\,  {f}_{n q s} \ .
\end{eqnarray}

It is likely that implementing other field redefinitions and algebraic manipulations will further simplify the action. It would be desirable to take this action to a minimal form.

\section{$\alpha'$-deformations of the moduli space} \label{SEC:Moduli}

In this section we use our knowledge of the first order corrections to half-maximal gauged  supergravity  to investigate the structure of the effective potential.  From a phenomenological point of view, the general setting of gauged supergravity offers interesting perspectives, such as the possibility to stabilize all moduli in a controlled manner or a mechanism of spontaneous supersymmetry breaking.

A nonzero extremal value of the scalar potential presents a possibility to explain a small positive  value of the cosmological constant, as required by  observational data. However,  an accelerating spacetime must violate the strong energy condition and the no-go theorem of \cite{maldanun} guarantees that such solutions cannot be obtained from only the lowest order terms in the supergravity action.
 The sub-leading corrections to the four dimensional scalar potential obtained in the previous section offer the possibility  not only to modify the Minkowski minima to a small value, but they could also stabilize some of the massless modes, modify the  flat directions of the lowest order theory or change the slow-roll behavior in inflationary models.
 Although four dimensional maximally symmetric de Sitter solutions have been ruled out in
the perturbative $\alpha'$ expansion of string theory from generic analysis of both the spacetime \cite{gau} and the worldsheet \cite{kleba} theories,  the  $\alpha'$-corrections can be combined with non-perturbative quantum corrections or localized sources
to produce solutions with properties that cannot be obtained from two-derivative supergravity. Actually,
there are examples of $AdS_4$ solutions at large internal volume  in type IIB string theory \cite{quevedo} or in the heterotic string \cite{anguelova}, in which the leading order Minkowski ground states are broken by higher-derivative terms that generate a nonzero cosmological constant.

With the motivation of better understanding  the effect of the $\alpha'$-corrections on the vacua of the zeroth order theory,  we  focus on the analysis of the effective potential (\ref{efpot}). It is important to stress that the $\alpha'$-corrections in (\ref{efpot}) cannot be eliminated by field redefinitions, as shown in Appendix \ref{App:Potential}. Moreover we emphasize that,
  since  all the terms in the heterotic effective action at string tree level scale uniformly with the dilaton, so does the four-dimensional effective scalar potential which does not depend on the dilaton otherwise. Hence the  dilaton equation of motion  implies either that the dilaton diverges or that the potential vanishes, and then   at lowest order in string perturbation theory the heterotic effective action can only lead to Minkowski solutions,  as shown  in \cite{gau}. As the dilaton  only appears as an overall multiplicative factor in the effective scalar potential,
 in the following analysis we will ignore this factor and restrict  attention to the rest of the moduli.
To be specific, we concentrate on the $\alpha'$-corrections to the Minkowski critical points of seven dimensional half-maximal supergravity with geometric gaugings \cite{Dibitetto:2015bia}.

 As we commented in the previous sections, the $\alpha'$-corrections to half-maximal supergravities originated from GSS reductions of DFT are those obtained by the choice of parameters $a=-\alpha',\;b=0$. Hence, the resulting scalar potential has the form
\begin{eqnarray}
U(\Phi)
&=&
U_0(\Phi)
+\alpha'\;U_1(\Phi)
+{\cal O}\left(\alpha'{}^2\right)\cr
&=&
e^{-2\phi} \left(V_0- \alpha'\, {\cal V}^{(-)}\right)+{\cal O}\left(\alpha'{}^2\right)\;,\,
\label{pot}
\end{eqnarray}
where $\Phi$ generically denotes the scalar fields, $V_0$ is given in (\ref{V0}) and ${\cal V}^{(-)}$  in (\ref{calV}).

To find the critical points of \eqref{pot} we have to solve the following equation:
\begin{eqnarray}
\partial_I U(\Phi_P)
=
\partial_I U_0(\Phi_P)
+\alpha'\;\partial_I U_1(\Phi_P)
+{\cal O}\left(\alpha'{}^2\right)
=
0
\, .
\label{eq:alphaminimum}
\end{eqnarray}
Solving order by order, we obtain the corrected position in the moduli space
\begin{eqnarray}
\Phi_P^I
=
\Phi_0^I
+\alpha' \Phi_1^I
+{\cal O}\left(\alpha'{}^2\right)
\, ,
\end{eqnarray}
where $\Phi_0^I$ denote the coordinates of a known critical point for $U_0(\Phi)$ and $\Phi_1^I $ is the shift generated by the first order corrections of the scalar potential. When Taylor expanding the terms in \eqref{eq:alphaminimum} and truncating the ${\cal O}\left(\alpha'{}^2\right)$ contributions, we have
\begin{eqnarray}
\partial_I U(\Phi_P)
=
\alpha'\;\left(\Phi_1^J\partial_J\partial_I U_0(\Phi_0)
+ \partial_I U_1(\Phi_0)\right)
+{\cal O}\left(\alpha'{}^2\right)=0
\, ,
\label{Pcrit}
\end{eqnarray}
where we have considered that the leading order is trivially satisfied. If the Hessian is invertible, the first $\alpha'$ order can be solved algebraically as
\begin{eqnarray}
\Phi_1^I =- \partial_J U_1(\Phi_0) \left(\partial_I\partial_J U_0(\Phi_0)\right)^{-1}
\, .
\label{Phi1}
\end{eqnarray}

On the other hand, when $\partial_J\partial_I U_0(\Phi_0)$ has vanishing determinant, the analysis is more subtle. In this case,
the Hessian has to be diagonalized, in order to separate a vanishing block, and thus in the non-vanishing directions one can invert it and find the corresponding $\Phi_1^J$ using (\ref{Phi1}).
In the directions in which the Hessian is null, the condition (\ref{Pcrit}) for $\Phi_P^I$ to be a critical point at order $\alpha'$  reduces to $\partial_I U_1(\Phi_0)=0$, which is a non trivial condition. Actually $\Phi_0$ is not a point when there are flat directions, then the condition $\partial_I U_1(\Phi_0)=0$ can either $(1)$ still have flat directions or $(2)$ completely fix $\Phi_0$ when the solution is unique or $(3)$ have no solution at all, which means that the critical point of the zeroth order theory disappears when $\alpha'$-corrections are turned on.

If the critical point does exist then there is a cosmological constant\footnote{$\Lambda$ only depends on $\Phi_0$ because the critical point condition eliminates the dependence on $\Phi_1$.} $\Lambda$,
\begin{eqnarray}
\Lambda
&=&
U(\Phi_P)
=
U_0(\Phi_0)+\alpha'\,U_1\left(\Phi_0\right)
+{\cal O}\left(\alpha'{}^2\right)
\, .
\end{eqnarray}

Let us now consider how this works in a particular example. For instance, we will explore here if the first order $\alpha'$-corrections affect the vacua structure of  half-maximal supergravity, with $n=7$, $d = 3$ and $N = 0$.

This theory possesses 16 supercharges and a global duality group $G_0={\mathbb R^+\times SO(3,3)}\approx \mathbb R^+\times SL(4)$. The linear constraints force the embedding tensor (ET) to transform in one of the following irreducible representations of $G_0$ in the $SL(4)$ branching (subindices stand for $\mathbb{R}^+$ weights)
\begin{equation}
\begin{array}{lclclclclc}
\Theta 	&  \in  	& \underbrace{\textbf{1}_{(-4)}}_{\theta} & \oplus & \underbrace{ \textbf{10}^{\prime}_{(+1)}}_{Q_{(ij)}} 	& \oplus 	& \underbrace{ \textbf{10}_{(+1)}}_{{\tilde{Q}}^{(ij)}}	& \oplus 	&	\underbrace{\textbf{6}_{(+1)}}_{\xi_{[ij]}} &.
\end{array}
\notag
\end{equation}

Only a subsector of the full set of available supersymmetric deformations is captured by GSS reductions of DFT. In particular, the gaugings considered here are such that  $\theta=\xi_{ij}=0$. Notice that the notation here is exactly the opposite to  the one in \cite{Dibitetto:2015bia}. Here we take indices $i,j,k$ to belong to the fundamental representation of $SL(4)$ while $m,n,p$ are  indices in the fundamental of $SO(3,3)$.
The deformations $Q$ and $\tilde Q$ can be easily related to the gaugings $f_{m n p}$ through the following expressions. First, we define
\begin{equation}
(X_{i_1 i_2})_{j_1 j_2}{}^{k_1 k_2}
=
\frac{1}{2}\,\delta^{[k_1}_{[i_1}\,Q_{i_2 ][     j_1}\,\delta^{ k_2] }_{ j_2 ]} \,
+\frac{1}{4} \, \epsilon_{li_1 i_2 [ j_1}\, \tilde{Q}^{l[k_1}\,\delta^{k_2]}_{j_2]}	
\, ,\label{SL4fluxes}
\end{equation}
Then, the fluxes $f_{mnp}$ in (\ref{V0}) are related to those in (\ref{SL4fluxes}) through the 't Hooft symbols $G_m$, which map the fundamental representation of $SO(3,3)$ into the anti-symmetric two-form of $SL(4)$ ,
\begin{equation}
f_{mnp} = [G_m]^{i_1i_2} \  [G_n]^{j_1j_2} \ [G_p]_{k_1k_2} \ (X_{i_1i_2})_{j_1j_2}{}^{k_1k_2}
\, .
\end{equation}
We note that when $\tilde Q^{(ij)} Q_{(i j)} = 0$, then $f_{m n p} f^{m n p} = 0$ and the gaugings satisfy the constraints of  maximal supergravity \cite{Aldazabal:2011yz}. From a GSS compactification point of view, this constraint holds for geometric reductions that satisfy the strong constraint \cite{Aldazabal:2011nj}.

We applied the approach described at the beginning of this section together with the {\it go to the origin} GTTO setting \cite{Dibitetto:2011gm} to two different sets of Minkowski vacua in $n = 7$ half-maximal supergravity, namely two 2-parameter families given by
\begin{equation}
Q = \text{diag}(\lambda, \lambda, 0, 0)\, , \qquad\qquad
\tilde Q = \text{diag}(0, 0, \mu, \mu)\, ,
\end{equation}
and
\begin{equation}
Q = \text{diag}(\lambda, \lambda, \mu, \mu)\, , \qquad\qquad
\tilde Q = \text{diag}(\mu, \mu,\lambda,\lambda)\, ,
\end{equation}
respectively, where $\lambda, \mu \in {\mathbb R}$. These vacua are solutions of the CSO(2,0,2) and the SO(2,2) gaugings, respectively. We refer to Table 4 of \cite{Dibitetto:2015bia} for more details. As they satisfy $\tilde Q^{(ij)} Q_{(i j)} = 0$,  both deformations are locally geometric and can be uplifted to the maximal theory \cite{Samtleben:2005bp}.

The result is that for these two cases, condition \eqref{Pcrit} is trivially satisfied. As these vacua already have flat directions at zeroth order, it means that  what we called condition (1)  above holds and so the position of the critical point remains unchanged. In addition $U_1(\Phi_0)=0$ in both cases, which means that the $\alpha'$-corrections to the scalar potential do not contribute to the cosmological constant, and the Minkowski vacua survive in both configurations.  Therefore we rule out in these particular cases the possibility of having a de Sitter vacuum upon considering $\alpha'$-corrections to the scalar potential, even when  ignoring the dilaton direction.

It would be interesting to push this investigation forward to understand if this is a generic behaviour. Having the $\alpha'$-corrected scalar potential of gauged supergravities, it is now possible to explore these issues in full generality. Not only corrections to the cosmological constant are worth studying, also corrections to massless scalar modes could drive lowest order vacua unstable (or stabilize it) or even rule out inflationary behaviour at lowest order. We hope to come back to these issues in the future.

\section{Outlook and concluding remarks} \label{SEC:Conclu}

The traditional  DFT is equipped with a duality covariant gauge symmetry principle based on a generalized Lie derivative that determines the two-derivative effective action uniquely \cite{Siegel:1993xq,Hull:2009mi}. Different parameterizations and choices of section allow to make contact with the standard universal bosonic sector of supergravity and lower-dimensional half-maximal gauged supergravities \cite{Aldazabal:2011nj}. Recently the duality covariant gauge symmetry principle was extended in the frame-formalism to include first-order deformations that account for the Green-Schwarz transformations of the heterotic string \cite{Marques:2015vua}. In addition, the deformations are in fact general enough to capture the first order corrections to the bosonic string as well as the $\alpha'$-geometry of the HSZ theory \cite{Hohm:2013jaa}.

Here we have revisited the generalized Green-Schwarz transformations and considered them from a broader perspective. Exploiting the fact that GSS compactifications  are effectively equivalent to gauging the theory \cite{Grana:2012rr}, we gauged the results in \cite{Marques:2015vua} and extended the parameterization of the generalized fields to include, in addition to the frame, two-form and dilaton, extra gauge and scalar fields. The freedom to choose  the dimensionality of the external and  internal spaces, the gauge group and the two free parameters that control the deformations permits to reach all the theories with this field content that enjoy an underlying $G$-duality symmetry, thus generalizing the results in \cite{Marques:2015vua}.

We have written the most general action in Section \ref{sec:The action}. Expressed in terms of generalized fluxes, this action includes the $26$-dimensional bosonic string, the $10$-dimensional heterotic string, and half-maximal supergravities in different dimensions, all corrected to first-order in $\alpha'$. While the first order corrections to the bosonic and heterotic strings are well known, and then constitute a validation of our results, the leading corrections to gauged supergravities had not  been computed before in full generality and are then a prediction of the formalism.

One of the most remarkable aspects of the effective action is that the scalar potential receives an unambiguous first order correction. Understanding how this deformation affects the vacuum structure is of interest, as flat directions in the moduli space could be lifted breaking the degeneracy of vacua with destabilized scalars, or changing the slow roll behavior in inflationary models.

Another promising line of research is to understand how to incorporate higher orders in this formalism. The Green-Schwarz transformations induce an infinite tower of $\alpha'$-corrections. The three-form field strength $\widehat H_{\mu \nu \rho}$ depends on the torsionful spin connection $\omega^{(-)}_{\mu a}{}^b$, the torsion being proportional to $\widehat H_{\mu \nu \rho}$ itself. This determines a system that can be worked out iteratively in an $\alpha'$ perturbative expansion. Second and higher-order corrections of this kind are not captured by the generalized Green-Schwarz transformations considered here because closure fails to hold at second-order in $\alpha'$. Finding a complete deformation that is exactly duality and gauge invariant is an open problem that deserves attention.

The parameter space can be further constrained by supersymmetry. We expect that only the deformations that correspond to the heterotic string $b = 0$ admit supersymmetrization, and it would be nice to check this explicitly. Even if an exactly closed form of  supersymmetric generalized Green-Schwarz transformations is found, constructing an exactly invariant action can have subtleties. We obtained here the first order corrections to the DFT generalized Ricci scalar. However, it is possible that unambiguous higher-derivative invariants exist that would trigger their own tower of $\alpha'$-corrections, leading for example to quartic Riemann terms and beyond. Understanding the full picture would be useful in order to have a complete classification of the constraints imposed by duality and supersymmetry.

Finally, other applications of this formalism arise: finding consistent higher-derivative deformations in Exceptional Field Theories or exploring if the generalized Green-Schwarz transformation, among others, can shed light on the discussion on large gauge transformations in DFT, etc. We hope to come back to these issues in the future.

~

\noindent {\bf \underline{Acknowledgments:}} JJF-M acknowledges support from JSPS Postdoctoral Fellowship and Fundaci\'on S\'eneca/Universidad de Murcia. D.M. thanks the organizers of the BIRS workshop {\it Double Field Theory, Exceptional Field Theory and their applications} for hospitality, while part of this work was being
completed. Support by A.S.ICTP, CONICET, UBA, ANPCyT and UNLP is also gratefully acknowledged.

~

\begin{appendix}
\section{Conventions and definitions}\label{SEC:Conventions}
In this appendix we introduce the notation used throughout the paper. Space-time and tangent space Lorentz indices are denoted $\mu, \nu, \dots$
and $a,b,\dots$, respectively. The internal double-Lorentz indices transformed by $H_i$ are denoted  $\alpha,\beta,\dots$ and internal indices rotated by global $G_i$ transformations are denoted $m,n,\dots$.
\subsection{Diffeomorphisms}
The Lie derivative of a tensor is given by
\be
L_{\xi} V_{\mu}{}^{\nu}= \xi^{\rho} \partial_{\rho}V_{\mu}{}^{\nu} + \partial_{\mu} \xi^{\rho} V_{\rho}{}^{\nu} - \partial_{\rho} \xi^{\nu} V_{\mu}{}^{\rho} \ .
\ee
The Christoffel connection is defined in terms of the metric as
\be
\Gamma_{\mu \nu}^{\rho} = \frac 1 2 g^{\rho \sigma} \left( \partial_{\mu} g_{\nu \sigma} + \partial_{\nu} g_{\mu \sigma} - \partial_{\sigma} g_{\mu \nu}\right) \ , \ \ \ \ \ \ \Gamma_{[\mu \nu]}^{\rho} = 0 \ ,
\ee
and transforms anomalously under infinitesimal diffeomorphisms (whenever the Lie derivative acts on a
non-tensorial object, we use the convention that it acts as if it
were covariant)
\be
\delta_{\xi} \Gamma_{\mu \nu}^\rho = L_{\xi}  \Gamma_{\mu \nu}^\rho + \partial_{\mu} \partial_{\nu} \xi^{\rho} \ ,
\ee
so it allows to define a covariant derivative, given by
\be
\nabla_{\rho} V_{\mu}{}^{\nu} = \partial_{\rho} V_{\mu}{}^{\nu} - \Gamma_{\rho \mu}^{\sigma} V_{\sigma}{}^{\nu} + \Gamma_{\rho \sigma}^{\nu} V_{\mu}{}^{\sigma} \ .
\ee
The commutator of two covariant derivatives
\be
\left[ \nabla_{\mu} ,\ \nabla_{\nu}\right] V_{\rho}{}^{\sigma} = - R^\delta{}_{\rho \mu \nu} \, V_\delta{}^\sigma + R^\sigma{}_{\delta \mu \nu} V_\rho{}^\delta \ , \label{Bi1}
\ee
is expressed in terms of the Riemann tensor
\be
R^{\rho}{}_{\sigma \mu \nu} = \partial_{\mu} \Gamma_{\nu \sigma}^{\rho} - \partial_{\nu} \Gamma_{\mu \sigma}^\rho + \Gamma_{\mu \delta}^\rho \Gamma_{\nu \sigma}^\delta - \Gamma_{\nu \delta}^\rho \Gamma_{\mu \sigma}^\delta \ , \label{curvedRiemann}
\ee
which  symmetries and Bianchi identities are
\be
R_{\rho \sigma \mu \nu} = g_{\rho \delta} R^{\delta}{}_{\sigma \mu \nu} = R_{([\rho \sigma][\mu \nu])}
\ , \ \ \ \
R^{\rho}{}_{[\sigma \mu \nu]} = 0 \ , \ \ \ \ \nabla_{[\mu} R_{\nu \lambda]}{}^\rho{}_\sigma = 0 \ .
\ee
Traces of the Riemann tensor give the Ricci tensor and scalar, respectively
\be
R_{\mu \nu} = R^{\rho}{}_{\mu \rho \nu} \ , \ \ \ \ \ \ R = g^{\mu \nu} R_{\mu \nu} \ .
\ee
\subsection{External Lorentz transformations}
The (inverse) metric can be written in terms of a (inverse) frame field
\be
g_{\mu \nu} = e_{\mu}{}^a g_{a b} e_{\nu}{}^b \ , \ \ \ \ \ g^{\mu \nu} = e_{a}{}^\mu g^{a b} e_{b}{}^\nu \ ,
\ee
where $g_{a b}$ is the Minkowski metric, and they satisfy the following identities
\be
e_a{}^\mu e_\mu{}^b = \delta^b_a \ , \ \ \ \ \ e_\mu{}^a e_a{}^\nu = \delta^\nu_\mu
\ , \ \ \ \ \ e_{a}{}^\mu = g^{\mu \nu} e_\nu{}^b g_{b a} \ .
\ee
Under Lorentz and infinitesimal diffeomorphism transformations, the frame
field changes as follows
\be
\delta e_{\mu}{}^a = L_\xi e_{\mu}{}^a + e_{\mu}{}^b \Lambda_b{}^a \ , \ \ \ \ \
\delta e_{a}{}^\mu = L_\xi e_{a}{}^\mu - \Lambda_a{}^b e_{b}{}^\mu
 \ , \ \ \ \ \
\Lambda_{a b} = \Lambda_a{}^c g_{c b} = - \Lambda_{b a} \ .
\ee
We also consider a spin connection defined in terms of the frame field
\be
\omega_{\mu a}{}^b = \partial_{\mu} e_{\nu}{}^b e_a{}^\nu - \Gamma_{\mu \nu}^\rho e_\rho{}^b e_a{}^\nu \ , \label{spinconnection}
\ee
that transforms as
\be
\delta \omega_{\mu a}{}^b = L_\xi \omega_{\mu a}{}^b + \partial_\mu \Lambda_{a}{}^b + \omega_{\mu a}{}^c \Lambda_c{}^b - \Lambda_a{}^c \omega_{\mu c}{}^b \ .
\ee
The Riemann tensor can also be written as an adjoint Lorentz-valued two-form,
expressed in terms of the spin connection as
\be
R_{\mu \nu a}{}^b = \partial_{\mu} \omega_{\nu a}{}^b - \partial_{\nu}
\omega_{\mu a}{}^b + \omega_{\mu a}{}^c \omega_{\nu c}{}^b  -   \omega_{\nu a}{}^c
\omega_{\mu c}{}^b \ . \label{twoformRiemann}
\ee
This form of the Riemann tensor transforms as
\be
\delta R_{\mu \nu a}{}^b = L_\xi R_{\mu \nu a}{}^b + R_{\mu \nu a}{}^c \Lambda_c{}^b -
\Lambda_a{}^c R_{\mu \nu c}{}^b \ ,
\ee
and is related to the  Riemann tensor (\ref{curvedRiemann}) through a
frame rotation
\be
R_{\mu \nu a}{}^b e_{b}{}^\rho e_\sigma{}^a = - R^{\rho}{}_{\sigma \mu \nu} \ .
\ee
The Chern-Simons three-form is defined as
\be
\Omega_{\mu \nu \rho} = \omega_{[\mu a}{}^b \partial_{\nu} \omega_{\rho] b}{}^a + \frac 2 3
\omega_{[\mu a}{}^b \omega_{\nu b}{}^c \omega_{\rho] c}{}^a \ ,
\ee
and it transforms under infinitesimal diffeomorphisms and Lorentz transformations as
\be
\delta \Omega_{\mu \nu \rho} = L_\xi \Omega_{\mu \nu \rho} +
\partial_{[\mu} \left(\omega_{\nu a}{}^b \partial_{\rho]} \Lambda_b{}^a\right) \ .
\ee
The Chern-Simons three-form satisfies the identity
\be
\nabla_{[\mu} \Omega_{\nu \rho \sigma]} = \frac 1 4 R_{[\mu \nu a}{}^b R_{\rho \sigma] b}{}^a \ .
\ee
\subsection{Gauge transformations}
Generic gauge tensors $T_m{}^n$ transform as follows
\be
\delta T_m{}^n = L_\xi T_m{}^n - f_{pm}{}^q \lambda^p T_q{}^n + f_{pq}{}^n \lambda^p T_m{}^q \ .
\ee
Their derivatives fail to transform tensorially, and then one has to introduce a covariant derivative
\be
\nabla_\mu T_m{}^n = \partial_\mu T_m{}^n + f_{pm}{}^q A_\mu{}^p T_q{}^n - f_{pq}{}^n A_\mu{}^p T_m{}^q \ ,
\ee
where the gauge connections $A_\mu{}^m$ transform as
\be
\delta A_\mu{}^m = L_\xi A_\mu{}^m + \partial_\mu \lambda^m + f_{pq}{}^m \lambda^p A_\mu{}^q \ .
\ee
The two-form curvature of the gauge fields
\be
F_{\mu \nu}{}^m = 2 \partial_{[\mu} A_{\nu]}{}^m - f_{pq}{}^m A_\mu{}^p A_\nu{}^q \ ,
\ee
 is a tensor both under diffeomorphisms and gauge transformations, and so it transforms covariantly
\be
\delta F_{\mu \nu}{}^m = L_\xi F_{\mu \nu}{}^m + f_{pq}{}^m \lambda^p F_{\mu \nu}{}^q \ .
\ee
Throughout the paper, whenever we write a covariant derivative acting on tensors with mixed indices, we assume that the derivative is covariant with respect to  both diffeomorphisms and gauge transformations. Then, for example we have
\be
\nabla_{\mu} F_{\nu \rho}{}^m = \partial_\mu F_{\nu \rho}{}^m - \Gamma_{\mu \nu}^\sigma F_{\sigma \rho}{}^m - \Gamma_{\mu \rho}^\sigma F_{\nu \sigma}{}^m - f_{pq}{}^m A_\mu{}^p F_{\nu \rho}{}^q \ ,
\ee
which in turn implies
\be
\nabla_{[\mu}F_{\nu \rho]}{}^m = 0 \, . \label{BIdF}
\ee
The commutator of two covariant derivatives acting on gauge tensors satisfies the identity
\be
[\nabla_\mu ,\, \nabla_\nu]T^m = f_{pq}{}^m T^p F_{\mu \nu}{}^q \ .
\ee
The scalar fields $M_{m n}$ are gauge tensors and diffeomorphism scalars, so they transform as follows
\be
\delta M_{m n} = L_\xi M_{m n} - 2 f_{p (m}{}^q  M_{n) q} \lambda^p \ .
\ee
Finally, we define the gauge Chern-Simons three-form as
\be
\Omega^{(g)}_{\mu \nu \rho} = A_{[\mu}{}^m \partial_\nu A_{\rho] m} - \frac 1 3 f_{m n p} A_\mu{}^m A_\nu{}^n A_\rho{}^p \ ,
\ee
which transforms as
\be
\delta \Omega^{(g)}_{\mu \nu \rho} = L_\xi \Omega^{(g)}_{\mu \nu \rho} + \partial_{[\mu} \left( A_\nu{}^m \partial_{\rho]} \lambda_m \right) \ .
\ee
This gauge Chern-Simons three-form satisfies the identity
\be
\nabla_{[\mu} \Omega^{(g)}_{\nu \rho \sigma]} = \frac 1 4 F_{[\mu \nu}{}^m F_{\rho \sigma]m} \ .
\ee
\subsection{Internal double-Lorentz transformations}
The internal Lorentz transformations are parameterized by infinitesimal parameters $\Lambda_{\alpha \beta}$ that leave the $H_i$ metrics invariant
\be
\delta \kappa_{\alpha \beta} = 2 \Lambda_{(\alpha \beta)} = 0 \ , \ \ \ \ \ \delta M_{\alpha \beta} = 2 M_{\gamma (\alpha} \Lambda^\gamma{}_{\beta)} = 0 \ .
\ee
This in turn implies that it has the following projections under (\ref{internalPs1})-(\ref{internalPs2})
\be
\Lambda_{\alpha \beta} = \Lambda_{\underline{\alpha} \underline{\beta}} + \Lambda_{\overline{\alpha} \overline{\beta}} \ , \ \ \ \ \Lambda_{\underline{\alpha} \overline{\beta}} = \Lambda_{\overline{\alpha} \underline{\beta}} = 0 \ . \label{LambdaSplit}
\ee
The only field with a non-trivial internal double-Lorentz transformation is the internal scalar frame $\Phi_m{}^\alpha$ (or its inverse $\Phi_\alpha{}^m = \kappa_{\alpha \beta} \kappa^{m n} \Phi_n{}^\beta$)
\bea
\delta \Phi_m{}^\alpha &=& L_\xi \Phi_m{}^\alpha  - f_{pm}{}^q \lambda^p \Phi_q{}^\alpha + \Phi_m{}^\beta \Lambda_\beta{}^\alpha \\
\delta \Phi_\alpha{}^m &=& L_\xi \Phi_\alpha{}^m + f_{pq}{}^m \lambda^p \Phi_\alpha{}^q - \Lambda_\alpha{}^\beta \Phi_\beta{}^m \ .
\eea
One can define an internal Lorentz connection
\be
\omega_{\mu \alpha}{}^\beta = \Phi_\alpha{}^m \nabla_\mu \Phi_m{}^\beta \ ,
\ee
that transforms as follows
\be
\delta \omega_{\mu \alpha}{}^\beta = L_\xi \omega_{\mu \alpha}{}^\beta + \partial_\mu \Lambda_\alpha{}^\beta + \omega_{\mu \alpha}{}^\gamma \Lambda_\gamma{}^\beta - \Lambda_\alpha{}^\gamma \omega_{\mu \gamma}{}^\beta \ .
\ee

Due to the splitting (\ref{LambdaSplit}), it follows that this connection separates in two independent connections
\be
\omega^{(-)}_{\mu \alpha \beta} = \omega_{\mu \underline{\alpha} \underline{\beta}} \ , \ \  \ \ \omega^{(+)}_{\mu \alpha \beta} = \omega_{\mu \overline{\alpha} \overline{\beta}}
\ee
which transform as
\bea
\delta \omega^{(-)}_{\mu \alpha}{}^\beta &=& L_\xi \omega^{(-)}_{\mu \alpha}{}^\beta + \partial_\mu \Lambda_{\underline{\alpha}}{}^{\underline{\beta}} + \omega^{(-)}_{\mu \alpha}{}^\gamma \Lambda_{\underline{\gamma}}{}^{\underline{\beta}} - \Lambda_{\underline{\alpha}}{}^{\underline{\gamma}} \omega^{(-)}_{\mu \gamma}{}^\beta  \, ,\\
\delta \omega^{(+)}_{\mu \alpha}{}^\beta &=& L_\xi \omega^{(+)}_{\mu \alpha}{}^\beta + \partial_\mu \Lambda_{\overline{\alpha}}{}^{\overline{\beta}} + \omega^{(+)}_{\mu \alpha}{}^\gamma \Lambda_{\overline{\gamma}}{}^{\overline{\beta}} - \Lambda_{\overline{\alpha}}{}^{\overline{\gamma}} \omega^{(+)}_{\mu \gamma}{}^\beta \ .
\eea

We can now define the internal Lorentz Chern-Simons three-forms
\be
\Omega^{(i,\pm)}_{\mu \nu \rho} = \omega^{(\pm)}_{[\mu \alpha}{}^\beta \nabla_\nu \omega^{(\pm)}_{\rho] \beta}{}^\alpha + \frac 2 3 \omega^{(\pm)}_{[\mu \alpha}{}^{\beta} \omega^{(\pm)}_{\nu \beta}{}^{\gamma} \omega^{(\pm)}_{\rho] \gamma}{}^{\alpha} \ ,
 \ee
 which transform as follows
 \be
 \delta \Omega^{(i,\pm)}_{\mu \nu \rho} = L_\xi \delta \Omega^{(i,\pm)}_{\mu \nu \rho} + \partial_{[\mu} \left( \omega^{(\pm)}_{\nu \alpha}{}^\beta \partial_{\rho]} \Lambda_\beta{}^\alpha\right) \ .
 \ee

 Finally, we define the projected scalar Riemann tensors
 \be
 R^{(\pm)}_{\mu \nu \alpha}{}^\beta = 2 \partial_{[\mu} \omega^{(\pm)}_{\nu] \alpha}{}^\beta + 2 \omega^{(\pm)}_{[\mu \alpha}{}^\gamma \omega^{(\pm)}_{\nu]\gamma}{}^\beta \ ,
 \ee
 that transforms as
 \be
 \delta R^{(\pm)}_{\mu \nu \alpha}{}^\beta = L_\xi R^{(\pm)}_{\mu \nu \alpha}{}^\beta +  R^{(\pm)}_{\mu \nu \alpha}{}^\gamma \Lambda_\gamma{}^\beta - \Lambda_\alpha{}^\gamma  R^{(\pm)}_{\mu \nu \gamma}{}^\beta \ ,
 \ee
 and in terms of which  the following identity holds
 \be
 \nabla_{[\mu} \Omega^{(i,\pm)}_{\nu \rho \sigma]} = \frac 1 4 R^{(\pm)}_{[\mu \nu \alpha}{}^\beta  R^{(\pm)}_{\rho \sigma] \beta}{}^\alpha \ .
 \ee
\subsection{Green-Schwarz and Chern-Simons}

To lowest order in $\alpha'$, the two-form transforms as follows
\be
\delta B_{\mu \nu} = L_\xi B_{\mu \nu} + 2 \partial_{[\mu} \xi_{\nu]} + A_{[\mu}{}^m \partial_{\nu]} \lambda_m \ .
\ee
The covariant (to lowest order in $\alpha'$) three-form curvature tensor is then given by
\be
H_{\mu \nu \rho} = 3 \partial_{[\mu} B_{\nu \rho]} - 3 \Omega^{(g)}_{\mu \nu \rho} \ .
\ee
It is invariant under gauge transformations parameterized by $\xi_\mu$ and $\lambda^m$, and transforms as a three-form under diffeomorphisms.

We can now define the spin connections with torsion
\be
\omega^{(\pm)}_{\mu a b} = \omega_{\mu a b} \pm \frac 1 2 H_{\mu a b} \ , \ \ \ \ \ \ H_{\mu a b} = H_{\mu \nu \rho} e_a{}^\nu e_b{}^\rho \ ,
\ee
Note that we do not include any $\alpha'$-correction in the torsion, as we are only interested in first-order corrections in this paper.
When the two-form Riemann tensor is supra-labeled with a sign, we use the convention that it is defined as in (\ref{twoformRiemann}) but
in terms of the spin connection with torsion
\be
R^{(\pm)}_{\mu \nu a}{}^b = \partial_{\mu} \omega^{(\pm)}_{\nu a}{}^b - \partial_{\nu} \omega^{(\pm)}_{\mu a}{}^b + \omega^{(\pm)}_{\mu a}{}^c \omega^{(\pm)}_{\nu c}{}^b  -   \omega^{(\pm)}_{\nu a}{}^c \omega^{(\pm)}_{\mu c}{}^b \ .
\ee
The supra-labeled
with a sign torsionful Chern-Simons three-form is accordingly
\be
\Omega^{(e, \pm)}_{\mu \nu \rho} = \omega^{(\pm)}_{[\mu a}{}^b
  \partial_{\nu} \omega^{(\pm)}_{\rho] b}{}^a + \frac 2 3 \omega^{(\pm)}_{[\mu a}{}^b \omega^{(\pm)}_{\nu b}{}^c \omega^{(\pm)}_{\rho] c}{}^a \ .
\ee
The transformations of the torsionful spin connection, Riemann tensor and Chern-Simons three-form are as follows
\bea
\delta \omega^{(\pm)}_{\mu a}{}^b &=& L_\xi \omega^{(\pm)}_{\mu a}{}^b + \partial_\mu \Lambda_{a}{}^b + \omega^{(\pm)}_{\mu a}{}^c \Lambda_c{}^b - \Lambda_a{}^c \omega^{(\pm)}_{\mu c}{}^b \, ,\\
\delta R^{(\pm)}_{\mu \nu a}{}^b &=& L_\xi R^{(\pm)}_{\mu \nu a}{}^b + R^{(\pm)}_{\mu \nu a}{}^c \Lambda_c{}^b - \Lambda_a{}^c R^{(\pm)}_{\mu \nu c}{}^b\, ,\\
\delta \Omega^{(e,\pm)}_{\mu \nu \rho} &=& L_\xi \Omega^{(e,\pm)}_{\mu \nu \rho} + \partial_{[\mu} \left( \omega^{(\pm)}_{\nu a}{}^b  \partial_{\rho]} \Lambda_b{}^a\right) \ .
\eea
The Lorentz Chern-Simons three-forms satisfy the identities
\be
\nabla_{[\mu} \Omega^{(e,\pm)}_{\nu \rho \sigma]} = \frac 1 4 R^{(\pm)}_{[\mu \nu a}{}^b R^{(\pm)}_{\rho \sigma] b}{}^a \ .
\ee

When first order $\alpha'$-corrections are turned on, the two-form field receives a deformation in its gauge transformations
\bea
\delta B_{\mu \nu} &=& L_\xi  B_{\mu \nu} + 2 \partial_{[\mu} \xi_{\nu]} +  A_{[\mu}{}^m \partial_{\nu]}\lambda_m - \frac 1 2 \left(a \omega^{(-)}_{[\mu}{}^{\alpha \beta} - b \omega^{(+)}_{[\mu}{}^{\alpha \beta}\right) \partial_{\nu]}\Lambda_{\alpha \beta} \nn \\
       && - \frac 1 2 \left(a \omega_{[\mu}^{(-)ab} - b \omega_{[\mu}^{(+)ab}\right) \partial_{\nu]}\Lambda_{ab} \ ,
\eea
that forces a correction in its three-form field strength
\be
\widehat H_{\mu \nu \rho} = H_{\mu \nu \rho} - \frac 3 2 a \Omega^{(e,-)}_{\mu \nu \rho} + \frac 3 2 b \Omega^{(e,+)}_{\mu \nu \rho} - \frac 3 2 a \Omega^{(i,-)}_{\mu \nu \rho} + \frac 3 2 b \Omega^{(i,+)}_{\mu \nu \rho} \ . \label{widehatH}
\ee
The Bianchi identity for this tensor is given by
\bea
\nabla_{[\mu} \widehat H_{\nu \rho \sigma]} &=& - \frac 3 4 F_{[\mu \nu}{}^m F_{\rho \sigma] m} - \frac 3 8 a R^{(-)}_{[\mu \nu a}{}^b R^{(-)}_{\rho \sigma] b}{}^a + \frac 3 8 b R^{(+)}_{[\mu \nu a}{}^b R^{(+)}_{\rho \sigma] b}{}^a \nn \\
&& - \frac 3 8 a R^{(-)}_{[\mu \nu \alpha}{}^\beta R^{(-)}_{\rho \sigma] \beta}{}^\alpha + \frac 3 8 b R^{(+)}_{[\mu \nu \alpha}{}^\beta R^{(+)}_{\rho \sigma] \beta}{}^\alpha \, .\label{BIdH}
\eea

\section{Lowest order action, EOMs and  field redefinitions} \label{SEC:Redefs}

Here we briefly review the zeroth order action of half-maximal gauged supergravities. The action is given by
\be
S = \int d^n X \sqrt{-g} e^{-2 \phi} {\cal L}_0 \ ,
\ee
where
\bea
{\cal L}_0 &=& R + 4 \nabla_\mu \nabla^\mu \phi - 4 \nabla_\mu \phi \nabla^\mu \phi - \frac 1 {12} H_{\mu \nu \rho} H^{\mu \nu \rho} \nn\\
&& - \frac 1 4 F_{\mu \nu}{}^m F^{\mu \nu n} M_{m n} + \frac 1 8 \nabla_\mu M_{mn} \nabla^\mu M^{mn} - V_0 \ ,
\eea
and the scalar potential is
\be
V_0 = \frac 1 {12} f_{m p}{}^r f_{n q}{}^s M^{mn} M^{pq} M_{r s} + \frac 1 4 f_{m p}{}^q f_{n q}{}^p M^{m n} + \frac 1 6 f_{mnp} f^{mnp} \ .
 \ee

Varying the action with respect to the fields gives, up to total derivatives,
\be
\delta S = \int d^n X \sqrt{-g} e^{-2 \phi} \left(\Delta g^{\mu \nu} \delta g_{\mu \nu} + \Delta \phi \delta \phi + \Delta B^{\mu \nu} \delta B_{\mu \nu} + \Delta A^\mu{}_m \delta A_\mu{}^m + \Delta M^{m n} \delta M_{mn}\right) \ , \nn
\ee
where
\bea
\Delta \phi &=& - 2 {\cal L}_0 \nn \\
\Delta g_{\mu \nu} &=& \frac 1 4 g_{\mu \nu} \Delta \phi + R_{\mu \nu} + 2 \nabla_\mu \nabla_\nu \phi - \frac 1 4 H_{\mu \rho \sigma} H_{\nu}{}^{\rho \sigma} \nn \\
&& - \frac 1 2 F_{\mu \rho}{}^m F_\nu{}^{\rho n} M_{m n} + \frac 1 8 \nabla_\mu M_{mn} \nabla_\nu M^{mn} \nn \\
\Delta B_{\mu \nu} &=& \frac 1 2 \nabla^\rho H_{\rho \mu \nu} - \nabla^\rho \phi H_{\rho \mu \nu}\label{FieldRed}\\
\Delta A_\mu{}^m &=& A_\nu{}^m \Delta B^\nu{}_\mu - 2 \nabla^\nu \phi F_{\nu \mu n} M^{n m} + \nabla^\nu \left( F_{\nu \mu n} M^{n m} \right) \nn \\ && + \frac 1 2 H_\mu{}^{\rho \sigma} F_{\rho \sigma}{}^m + \frac 1 2 f_p{}^{q m} M_{q r} \nabla_\mu M^{r p} \nn\\
\Delta M_{mn} &=& \left(P_{m}{}^p \bar P_n{}^q + \bar P_m{}^p P_n{}^q \right) \left(
 - \frac 1 4 F_{\mu \nu p} F^{\mu \nu}{}_q + \frac 1 2 \nabla^\mu \phi \nabla_\mu M_{p q} - \frac 1 4 \nabla_\mu \nabla^\mu M_{p q} \right.\nn \\
&& \left. - \frac 1 4 f_{p u r} f_{q v s} M^{u v} M^{rs} - \frac 1 4 f_{p r}{}^s f_{q s}{}^r \right)\ ,\nn
\eea
with
\be
P_{m n} = \frac 1 2 \left(\kappa_{mn} - M_{mn}\right) \ , \ \ \ \bar P_{m n} = \frac 1 2 \left(\kappa_{mn} + M_{mn}\right)
\ee

Notice that under field redefinitions
\bea
&&g_{\mu \nu} \to g_{\mu \nu} + Dg_{(\mu \nu)} \ , \ \ \ B_{\mu \nu} \to B_{\mu \nu} + DB_{[\mu \nu]} \ , \ \ \ \phi \to \phi + D\phi \ , \nn \\
&& A_\mu{}^m \to A_\mu{}^m + DA_\mu{}^m \ , \ \ \ M_{m n} \to M_{m n} + DM_{(\underline{m}\overline{n})}\ ,
\eea
the lowest order action shifts (up to integration by parts) as
\bea
D S &=& \int d^n X \sqrt{-g} e^{-2 \phi} \left(\Delta g^{\mu \nu} Dg_{\mu \nu} + \Delta \phi D\phi + \Delta B^{\mu \nu} DB_{\mu \nu} \right.\nn \\
&& \quad \quad \quad \quad \quad \quad \quad \ \  \left.+ \Delta A^\mu{}_m D A_\mu{}^m + \Delta M^{m n} D M_{mn}\right) \ . \nn
\eea
Therefore, using equations of motion to simplify or eliminate terms in the first order part of the action, simply amounts to performing covariant first order field redefinitions, and is then a valid operation.

\section{Some technical details }
\subsection{Simplifying the action}\label{AppSF}

The purpose of this appendix is to give some explicit details in the derivation of (\ref{LeffSF}). The expressions of $L^{(\pm)}$, introduced in Section \ref{sec:The action} and related to ${\cal L}^{(\pm)}$ via (\ref{leff}) are our starting point. After some algebraic manipulations we obtain
\begin{eqnarray}
{\mathcal L}^{(\pm)}&=& - V^{(\pm)}\mp \frac{1}{4}\, {H}^{\mu \nu \rho} \Omega^{(\pm)}{}_{\mu \nu \rho}
+ \frac{1}{8}\, {R^{(\pm)}}^{\mu \nu \rho \sigma} {R^{(\pm)}}_{\mu \nu \rho \sigma}
\pm \frac14\, P^{(\pm)\,m n} F^{\mu}{}_{\rho m} F^{\nu \rho}{}_{n} R_{\mu \nu}\cr
&\mp& \frac{1}{4}\, {F}^{\mu \nu}\,_{m} {F}^{\rho \sigma}\,_{n} {R}_{\mu \nu \rho \sigma} {P^{(\mp)}}^{m n}
\pm \frac{1}{4}\, {F}^{\mu \nu}\,_{m} {F}^{\rho \sigma}\,_{n} {R}_{\mu \rho \nu \sigma} {P^{(\pm)}}^{m n}\cr
&-& \frac{1}{64}\, {\nabla}_{\mu}{{M}_{n p}}\,  {\nabla}_{\nu}{{M}_{q r}}\,  {\nabla}^{\mu}{{M}^{p r}}\,  {\nabla}^{\nu}{{M}^{m q}}\,
 (4\, {\eta}_{m}\,^{n} \mp {M}_{m}\,^{n})
+ \frac{1}{16}\, {\nabla}_{(\mu}{{\nabla}_{\nu)}{{M}_{m n}}\, }\,  {\nabla}^{\mu}{{\nabla}^{\nu}{{M}^{m n}}\, }\,   \cr
&-& \frac{1}{256}\, {\nabla}_{\mu}{{M}_{m n}}\,  {\nabla}_{\nu}{{M}_{p q}}\,  {\nabla}^{\mu}{{M}^{p q}}\,  {\nabla}^{\nu}{{M}^{m n}}\,
- \frac{1}{128}\, {\nabla}_{\mu}{{M}_{m n}}\,  {\nabla}_{\nu}{{M}^{m p}}\,  {\nabla}^{\mu}{{M}_{p q}}\,  {\nabla}^{\nu}{{M}^{n q}}\,  \cr
&\pm& \frac{1}{8} {f}_{m n p} {f}_{q r s} {\nabla}_{\mu}{{M}^{m1 p}}\,  {\nabla}^{\mu}{{M}^{m2 s}}\,  \left(
- {P^{(\mp)}}^{m q} ({\eta}^{r}\,_{m1} {\eta}^{n}\,_{m2} + 2\, {P^{(+)}}^{r}\,_{m1} {P^{(-)}}^{n}\,_{m2})\right. \cr
&&\;\;\;+\; {P^{(\pm)}}^{m q}
\left( 2\, {P^{(+)}}^{n}\,_{m1} {P^{(-)}}^{r}\,_{m2} + 2\, {\eta}^{r}\,_{m1} {\eta}^{n}\,_{m2}
 -{M}^{r n} {M}_{m1 m2}\right.\cr
&&\left.\left.\;\;\;\;\;\;\;\;\;\;\;\;\;\;\;\;\;\;\;\;\;\;\;\;\;\;\;\;\;\;\;\;\;\;\;\;\;\;\;\;\;\;\;\;\;\;\;\;\;\;\;\;\;\;\;\;\;\;\;\;\;\;
\;\;\;\;\;\;\;\;\;\;\;\;\;
  +\;{P^{(\mp)}}^{r n} ( {\eta}_{m2 m1} \pm 2\, {M}_{m2 m1}) \right) \right) \cr
&-& \frac{1}{32} {F}_{\mu \nu m} {F}_{\rho \sigma n} {F}^{\mu \nu}\,_{p} {F}^{\rho \sigma}\,_{q}
\left(M^{m n} M^{p q}+{\eta}^{m n} {\eta}^{p q} \right) \cr
&-&\frac{1}{64} {F}_{\mu \nu m} {F}_{\rho \sigma n} {F}^{\mu \rho}\,_{p} {F}^{\nu \sigma}\,_{q}
(M^{m n} M^{p q} \mp 2\, M^{m n} \eta^{p q}-7 {\eta}^{m n} {\eta}^{p q} \pm 4\left(\eta^{m p}\pm 2\, {M}^{m p} \right){M}^{n q} ) \cr
&\mp& \frac{1}{8}\, (2\, {P^{(\pm)}}^{m n} - {P^{(\mp)}}^{m n}) {\nabla}_{\mu}{{F}_{\nu \rho m}}\,  {\nabla}^{\mu}{{F}^{\nu \rho}\,_{n}}\cr
&\pm&\frac{1}{4} {P^{(\pm)}}^{m n} ({\nabla}^{\mu}{{F}_{\mu \nu m}}\,  {\nabla}_{\rho}{{F}^{\nu \rho}\,_{n}}\,
+ {\nabla}^{\mu}{{\nabla}_{\nu}{{F}_{\rho \mu n}}\, }\,  {F}^{\nu \rho}\,_{m}
+  {\nabla}^{\mu}{{\nabla}_{\nu}{{F}^{\rho \nu}\,_{n}}\, }\,  {F}_{\mu \rho m}) \cr
&-&\frac{1}{16}\, {F}^{\mu \rho}\,_{m} {F}_{\mu \nu n} {\nabla}_{\rho}{{\nabla}^{\nu}{{M}_{p q}}\, }\,
\left(M^{m p} M^{n q}-\eta^{m p} \eta^{n q}\right) \cr
&\pm& \frac{1}{64}\, {F}^{\mu \rho}\,_{m} {F}_{\mu \nu n} {\nabla}_{\rho}{{M}_{p q}}\,  {\nabla}^{\nu}{{M}^{p q}}\,
\left( \eta^{m n}\pm 3\,{M}^{m n} \right)\cr
&\mp& \frac{1}{32}{F}_{\mu \nu m} {F}^{\mu \nu}\,_{n} {\nabla}^{\rho}{{M}^{m p}}\,  {\nabla}_{\rho}{{M}^{n q}}\,
( {\eta}_{p q} \pm 4\, {M}_{p q}) \cr
&\pm&\frac{1}{8}\, {F}^{\mu \rho}\,_{m} {F}_{\mu \nu n} {\nabla}_{\rho}{{M}^{m p}}\,  {\nabla}^{\nu}{{M}^{n q}}\,
(2\, {P^{(\mp)}}_{p q} -  {P^{(\pm)}}_{p q}) \cr
&+& \frac{1}{4}\,{F}^{\mu \nu}\,_{m} {\nabla}_{\mu}{{F}_{\nu \rho n}}\,  {\nabla}^{\rho}{{M}^{m p}}\,
(2\, {P^{(\mp)}}^{n}\,_{p} +  {P^{(\pm)}}^{n}\,_{p})\cr
&-& \frac{1}{8} {F}^{\mu \nu}\,_{m} {\nabla}_{\rho}{{F}_{\mu \nu n}}\,  {\nabla}^{\rho}{{M}^{m p}}\,
( 2\, {P^{(\pm)}}^{n}\,_{p} + {P^{(\mp)}}^{n}\,_{p})
 - \frac{1}{4}\, {F}^{\mu \nu}\,_{m} {\nabla}_{\nu}{{M}^{m n}}\,  {\nabla}^{\rho}{{F}_{\mu \rho n}}\,  \cr
&\pm &
P^{(\pm)\,m n} \nabla^{\mu}F_{\mu \rho m} F^{\nu \rho}{}_{n} \nabla_{\nu}\phi
\pm P^{(\pm)\,m n} F_{\mu \rho m} \nabla^{\mu} F^{\nu \rho}{}_{n} \nabla_{\nu}\phi \cr
&\pm& P^{(\pm)\,m n} F_{\mu \rho m} F^{\nu \rho}{}_{n} \nabla^{\mu}\nabla_{\nu}\phi
+\frac12  F_{\mu \rho m} F^{\nu \rho}{}_{n} \nabla_{\nu}\phi \nabla^{\mu} M^{m n}
\mp P^{(\pm)\,m n} F_{\mu \rho m} F^{\nu \rho}{}_{n} \nabla_{\nu}\phi \nabla^{\mu}\phi
\cr
&+& \frac{1}{16}\, {f}_{m2 p r} {f}_{n q s} {F}_{\mu \nu m} {F}^{\mu \nu}\,_{m1}  {P^{(\pm)}}^{r s}
\left(\pm 2\, {M}^{m m2} ({\eta}^{n m1} {\eta}^{p q} +{M}^{n m1} {M}^{p q}) \right. \cr
&&\;\;\;\;\;\;\;\;\;\;\;\;\;\;\;\;\;\;\;\;\;\;\;\;\;\;\;\;\;\;\;\;\;\;\;\;\;\;\;\;\;\;\;\;\;\;\;\;\;\;\;\;\;\;\;\;\;\;\;\;\;\;\;\;\;\;
-\; \left.{\eta}^{p q} ({\eta}^{n m1} {\eta}^{m m2} +{M}^{n m1} {M}^{m m2}) \right) \cr
&-& \frac{1}{48}\, {F}^{\mu}\,_{\gamma m} {F}_{\mu \nu n} {F}^{\gamma \nu}\,_{p} {f}_{q r s}
\left( \, {M}^{m q} ({\eta}^{n r} {\eta}^{p s} + {M}^{n r} {M}^{p s})
+ 2(4\, {M}^{m q} \pm {\eta}^{m q}) {\eta}^{n r} {\eta}^{p s}\right) \cr
&+& \frac{1}{4}\, {F}_{\mu \nu m} {\nabla}^{\mu}{{M}_{p r}}\,  {\nabla}^{\nu}{{M}_{q s}}\,  {f}_{n}\,^{r s}
\left(\pm  {M}^{n [m} {P^{(\mp)}}^{p] q} + {\eta}^{m n} {\eta}^{p q} - {M}^{m p} {M}^{n q}\right)\cr
&-& \frac{1}{16}\, {F}_{\mu \nu m} {H}^{\mu \nu \rho} {M}^{m n} {M}^{p q} {\nabla}_{\rho}{{M}_{p}\,^{s}}\,  {f}_{n q s} \cr
&+& \frac{1}{64}\, {H}^{\mu \nu \rho} {H}_{\mu \nu \sigma} {\nabla}_{\rho}{{M}_{m n}}\,  {\nabla}^{\sigma}{{M}^{m n}}\,
\mp \frac{1}{32}\, {F}^{\mu \rho}\,_{n} {F}_{\mu \nu m} {H}^{\nu \sigma \gamma} {H}_{\rho \sigma \gamma}
\left(\eta^{m n}\pm 3\,{M}^{m n} \right)\cr
&\pm& \frac{1}{16}\, {F}^{\mu \nu}\,_{m} {F}^{\rho \sigma}\,_{n}
\left( {H}_{\mu \rho}\,^{\gamma} {H}_{\nu \sigma \gamma} {P^{(\mp)}}^{m n}
- {H}_{\mu \nu}\,^{\gamma} {H}_{\rho \sigma \gamma} {P^{(\pm)}}^{m n}\right) \cr
&\mp& \frac{1}{8}\, {F}^{\mu \nu}\,_{m} {F}^{\rho \sigma}\,_{n} {H}_{\mu \nu \rho}
\left(2\, {P^{(\mp)}}^{m p} - {P^{(\pm)}}^{m p}\right) {\nabla}_{\sigma}{{M}^{n}\,_{p}}\,
+ \frac{1}{16}\, {F}_{\mu \nu m} {F}^{\mu \rho}\,_{n} {H}^{\nu \sigma}\,_{\rho} {M}^{m p} {\nabla}_{\sigma}{{M}^{n}\,_{p}}\cr
&-& \frac{1}{4}\, {F}_{\mu \nu m} {H}^{\mu \rho \sigma} {\nabla}^{\nu}{{F}_{\rho \sigma n}}\,  {P^{(\pm)}}^{m n}
+ \frac{1}{4}\, {F}^{\rho}\,_{\mu m} {H}^{\mu \nu \sigma} {\nabla}_{\nu}{{F}_{\rho \sigma n}}\,  {P^{(\mp)}}^{m n}
- \frac14 \, F_{\mu \nu m} F_{\rho \sigma}{}^{m} \nabla^{\mu}H^{\nu \rho \sigma}\,,\cr&&\label{leff2}
\end{eqnarray}
Now notice that most of the terms containing derivatives of the dilaton can be rewritten as a total derivative, e.g.
\begin{eqnarray}
&&
e^{-2\phi}\left(\pm P^{(\pm)\,m n} \nabla^{\mu}F_{\mu \rho m} F^{\nu \rho}{}_{n} \nabla_{\nu}\phi
\pm P^{(\pm)\,m n} F_{\mu \rho m} \nabla^{\mu} F^{\nu \rho}{}_{n} \nabla_{\nu}\phi \right.\cr
&&\pm P^{(\pm)\,m n} F_{\mu \rho m} F^{\nu \rho}{}_{n} \nabla^{\mu}\nabla_{\nu}\phi
+\frac12  F_{\mu \rho m} F^{\nu \rho}{}_{n} \nabla_{\nu}\phi \nabla^{\mu} M^{m n}
\mp P^{(\pm)\,m n} F_{\mu \rho m} F^{\nu \rho}{}_{n} \nabla_{\nu}\phi \nabla^{\mu}\phi\left.\right)\cr
&&=\nabla^{\mu}\left(e^{-2\phi}\;P^{(\pm)\,m n} F_{\mu \rho m} F^{\nu \rho}{}_{n} \nabla_{\nu}\phi \right)
\pm e^{-2\phi} P^{(\pm)\,m n} F_{\mu \rho m} F^{\nu \rho}{}_{n} \nabla_{\nu}\phi \nabla^{\mu}\phi.
\label{DilatonTerms}
\end{eqnarray}
We can also perform some field redefinitions in order to get some extra simplifications (see Appendix \ref{SEC:Redefs}). In particular if we choose
\begin{eqnarray}
\delta g^{\mu\nu}&=&a\,\delta g^{\mu\nu}_{-}+b\,\delta g^{\mu\nu}_{+}\;,\;\;\;\;\;\;\;\;\;\;\;\;\;\;
\delta g^{\mu\nu}_{(\pm)}= \mp\frac14\, P^{(\pm) m n} F^{\mu\rho}{}_{m} F^{\nu}{}_{\rho n}\;,\cr
\delta \phi&=&a\,\delta \phi_{-}+b\,\delta \phi_{+}\;,\;\;\;\;\;\;\;\;\;\;\;\;\;\;\;
\delta \phi_{(\pm)}= \pm\frac{1}{16}\, P^{(\pm) m n} F^{\mu\nu}{}_{m} F_{\mu\nu n}\;,
\end{eqnarray}
we find that some terms in ${\mathcal L}^{(\pm)}$ are cancelled, in particular the one containing the Ricci tensor and the only term which is not a total derivative in the last line of (\ref{DilatonTerms}). It would be nice to explore if redefinitions of the scalars, the two-form and the gauge fields simplify the action further.

On the other hand we can use the Bianchi identity of the Riemann tensor $R_{\mu[\nu\rho\sigma]}=0$ to rewrite the terms
\begin{eqnarray}
&&\mp \frac{1}{4}\, {F}^{\mu \nu}\,_{m} {F}^{\rho \sigma}\,_{n} {R}_{\mu \nu \rho \sigma} {P^{(\mp)}}^{m n}
\pm \frac{1}{4}\, {F}^{\mu \nu}\,_{m} {F}^{\rho \sigma}\,_{n} {R}_{\mu \rho \nu \sigma} {P^{(\pm)}}^{m n}\cr
&&=\pm \frac{1}{8}\, {F}^{\mu \nu}\,_{m} {F}^{\rho \sigma}\,_{n} {R}_{\mu \nu \rho \sigma}
\left({P^{(\pm)}}^{m n}-2\,{P^{(\mp)}}^{m n}\right)\,,
\end{eqnarray}
which can then be absorbed into
\begin{eqnarray}
\frac18\,\tilde{R}{}^{(\pm)}_{\mu \nu \rho \sigma}\tilde{R}{}^{(\pm)\,\mu \nu \rho \sigma}&=&
\frac18\, R^{(\pm)}_{\mu \nu \rho \sigma} R^{(\pm)\,\mu \nu \rho \sigma}
\pm \frac18 F^{\mu \nu}{}_{m} F^{\rho \sigma}{}_{n} R_{\mu \nu \rho \sigma} \left(P^{(\pm)}{}^{m n}-2\, P^{(\mp)}{}^{m n}\right)\cr
&\mp&\frac{1}{16} F_{\mu \nu m} F_{\rho \sigma n} H^{\mu \rho \lambda}H^{\nu \sigma}{}_{\lambda}
\left(P^{(\pm)}{}^{m n}-2\, P^{(\mp)}{}^{m n}\right)\cr
&+&\frac{1}{32} F_{\mu \nu}{}^{m} F_{\rho \sigma}{}^{n} F^{\mu \nu p} F^{\rho \sigma q}
\left(2\left(P^{(+)}_{m n} P^{(+)}_{p q} + P^{(-)}_{m n} P^{(-)}_{p q} \right) - 5 P^{(+)}_{m n} P^{(-)}_{p q}\right)\cr
&+&\frac{1}{16} F_{\mu \nu}{}^{m} F_{\rho \sigma}{}^{n} F^{\mu \rho p} F^{\nu \sigma q}
\left(2 P^{(\mp)}_{m n} P^{(\mp)}_{p q} - P^{(\pm)}_{m n} P^{(\pm)}_{p q} + P^{(+)}_{m n} P^{(-)}_{p q}\right),\;\;\;\;\;\;\;\;
\label{RtildeSq}
\end{eqnarray}
where we have used the Bianchi identity of the three-form (\ref{BIdH}), in order to rewrite the term containing $\nabla_{\mu} H_{\nu\rho \sigma}$ in $\tilde{R}{}^{(\pm)}_{\mu \nu \rho \sigma}\tilde{R}{}^{(\pm)\,\mu \nu \rho \sigma}$ as
\begin{eqnarray}
\frac18\,F^{\mu \nu m} F^{\rho \sigma n} \nabla_{\mu} H_{\nu\rho \sigma} (P^{(\pm)}_{m n} -2 P^{(\mp)}_{m n})=
\frac18\,F^{\mu \nu m} F^{\rho \sigma n} \nabla_{[\mu} H_{\nu\rho \sigma]} (P^{(\pm)}_{m n} -2 P^{(\mp)}_{m n})
\;\;\;\;\;\;\;\;\;\;\;\;\;\;\;\nonumber\\
\;\;\;\;\;\;\;
= -\frac{1}{32} F^{\mu \nu m} F^{\rho \sigma n} \left( F_{\mu \nu}{}^{p} F_{\rho \sigma p} -2 F_{\mu \rho}{}^{p} F_{\nu \sigma p} \right)
(P^{(\pm)}_{m n} -2 P^{(\mp)}_{m n}).\;\;\;\;\;\;\;
\end{eqnarray}
This condition can also be used to put the last term in ${\mathcal L}^{\pm}$ in the form
\begin{eqnarray}
-\frac14 F^{\mu \nu m} F^{\rho \sigma}{}_{m} \nabla_{\mu}H_{\nu \rho \sigma}
= \frac{1}{16} F_{\mu \nu m} F_{\rho \sigma}{}^{m}
\left(F^{\mu \nu n} F^{\rho \sigma}{}_{n}-2 F^{\mu \rho n} F^{\nu \sigma}{}_{n}\right)\,.
\end{eqnarray}

On the other hand we can use the Bianchi identity for the field strength (\ref{BIdF}),
which implies ${\nabla}_{\rho}{{F}_{\mu \nu m}}\,  {\nabla}^{\rho}{{F}^{\mu \nu}\,_{n}}=-2\,{\nabla}_{\rho}{{F}_{\mu \nu m}}\, {\nabla}^{\mu}{{F}^{\nu \rho}\,_{n}}$ to rewrite
\begin{eqnarray}
&&\pm \frac{1}{4} {P^{(\pm)}}^{m n} ({\nabla}^{\mu}{{F}_{\mu \nu m}}\,  {\nabla}_{\rho}{{F}^{\nu \rho}\,_{n}}\,
+ {\nabla}^{\mu}{{\nabla}_{\nu}{{F}_{\rho \mu n}}\, }\,  {F}^{\nu \rho}\,_{m}
+  {\nabla}^{\mu}{{\nabla}_{\nu}{{F}^{\rho \nu}\,_{n}}\, }\,  {F}_{\mu \rho m})
\;\;\;\;\;\;\;\;\;\;\;\;\;\;\;\;\;\;\;\;\;\;\;\;\;\cr
&&\mp \frac{1}{8}\, (2\, {P^{(\pm)}}^{m n} - {P^{(\mp)}}^{m n}) {\nabla}_{\mu}{{F}_{\nu \rho m}}\,  {\nabla}^{\mu}{{F}^{\nu \rho}\,_{n}}
\;\;\;\;\;\;\;\;\;\;\;\;\;\;\;\;\;\;\;\;\;\;\;\;\;\;\;\;\;\;\;\;\;\;\;\;\;\;\;\;\;\;\;\;\;\;\;\;\;\;\cr
&&\;\;\;\;\;\;\;\;\;\;\;\;\;\;\;\;\;\;\;\;\;\;\;\;\;
=  \pm \frac{1}{4} {P^{(\pm)}}^{m n} {\nabla}^{\mu}{\nabla}_{\rho}\left({F}_{\mu \nu m} {F}^{\nu \rho}\,_{n}\right)
- \frac{1}{8}\, M^{m n} {\nabla}_{\mu}{{F}_{\nu \rho m}}\,  {\nabla}^{\mu}{{F}^{\nu \rho}\,_{n}}.
\end{eqnarray}
Again by using (\ref{BIdF}) we can write
\begin{eqnarray}
&&- \frac{1}{4}\, {F}_{\mu \nu m} {H}^{\mu \rho \sigma} {\nabla}^{\nu}{{F}_{\rho \sigma n}}\,  {P^{(\pm)}}^{m n}
+ \frac{1}{4}\, {F}^{\rho}\,_{\mu m} {H}^{\mu \nu \sigma} {\nabla}_{\nu}{{F}_{\rho \sigma n}}\,  {P^{(\mp)}}^{m n}
\;\;\;\;\;\;\;\;\;\;\;\;\;\;\;\;\;\;\;\;\;\;\;\;\;\;\;\;\;\;\;\;\;\;\;\;\;\;\cr
&&\;\;\;\;\;\;\;\;\;\;\;\;\;\;\;\;\;\;\;\;\;\;\;\;\;\;\;\;\;\;\;\;\;\;\;\;\;\;\;\;\;\;\;\;\;\;\;\;\;\;
= - \frac{1}{8}\, {F}_{\mu \nu m} {H}^{\mu \rho \sigma} {\nabla}^{\nu}{{F}_{\rho \sigma n}} \,
\left({P^{(\mp)}}^{m n}+2\,{P^{(\pm)}}^{m n}\right),
\end{eqnarray}
 and
\begin{eqnarray}
\frac{1}{4}\,{F}_{\mu \nu m} {\nabla}^{\mu}{{F}^{\nu \rho}\,_{n}}\,  {\nabla}_{\rho}{{M}^{m p}}\,
(2\, {P^{(\mp)}}^{n}\,_{p} +  {P^{(\pm)}}^{n}\,_{p})
- \frac{1}{8} {F}_{\mu \nu m} {\nabla}_{\rho}{{F}^{\mu \nu}\,_{n}}\,  {\nabla}^{\rho}{{M}^{m p}}\,
( 2\, {P^{(\pm)}}^{n}\,_{p} + {P^{(\mp)}}^{n}\,_{p})\nonumber\\
\;\;\;\;\;\;=\;\;
 -\frac{3}{8}\,{F}_{\mu \nu m} {\nabla}_{\rho}{{F}^{\mu \nu}\,_{n}}\,  {\nabla}^{\rho}{{M}^{m n}}\;.
\;\;\;\;\;\;\;\;\;\;\;\;\;\;\;\;\;\;\;\;\;\;\;\;\;\;\;\;\;\;\;\;\;\;\;\;\;\;\;
\end{eqnarray}
Hence plugging all these equations into (\ref{leff2}), we obtain  (\ref{LeffSF}).

\subsection{The $\alpha'$-corrections to the scalar potential}\label{App:Potential}
The purpose of this appendix is to show that the $\alpha'$-corrections to the scalar potential cannot be eliminated through field redefinitions. Let us discuss, without loss of generality the case $b = 0$. If such redefinitions existed, then $V_\alpha = -  \alpha' {\cal V}^{(-)}$ in (\ref{calV}) should be reproduced by the scalar part of $\left.\Delta M^{mn}\delta M_{mn}\right|_{scalar}=-\delta^{mn} V_0\delta M_{mn}$ (see (\ref{FieldRed})), where
 \begin{eqnarray}
\delta^{mn}V_0&=&-2\,S_{m'n'} \left(P^{(+)mm'}P^{(-)nn'}+P^{(-)mm'}P^{(+)nn'}\right)
               =-2\,\left(S^{\overline m \underline n} + S^{\underline m \overline n} \right)\cr
S_{m'n'}&=&f_{m'pq} f_{n'p'q'} P^{(+)pp'} P^{(-)qq'}
\end{eqnarray}
The first line of $V_{\alpha}$ in (\ref{calV})  can be rewritten as
\begin{eqnarray}
&&\left( P^{(-)}_{m m'} P^{(-)}_{n n'} P^{(+)}_{p p'}
					- P^{(+)}_{m m'} P^{(+)}_{n n'} P^{(-)}_{p p'} \right)
					P^{(-)}_{q q'} P^{(-)}_{r r'} P^{(+)}_{s s'}
					 f^{m p q} f^{n p' q'} f^{m' r s} f^{n'r's'}\;=\;\;\;\;\;\;\;\;\;\;\;\;\;\;\;\;\;\;\;\;\cr
					&&\;\;\;\;\;\;\;\;\;\;\;\;\;\;\;\;\;\;\;\;\;\;\;\;\;\;\;\;\;\;=\; S^{m' n'} \left( P^{(-)}_{m m'} P^{(-)}_{n n'} P^{(+)}_{p p'}
					- P^{(+)}_{m m'} P^{(+)}_{n n'} P^{(-)}_{p p'} \right)
					P^{(-)}_{q q'} f^{m p q} f^{n p' q'} \cr
					&&\;\;\;\;\;\;\;\;\;\;\;\;\;\;\;\;\;\;\;\;\;\;\;\;\;\;\;\;\;\;=\;
					\left(  S^{\underline m \underline n} P^{(+)}_{p p'}
					- S^{\overline m \overline n} P^{(-)}_{p p'} \right)
					P^{(-)}_{q q'} f^{m p q} f^{n p' q'} \, ,
\end{eqnarray}
and so we see that the projected components of $S$ above do not agree with those in $\delta^{mn}V_0$.

The situation is even worse for the second line of (\ref{calV}), as it is not  possible to generate the $S_{mn}$ factor. Indeed to generate it, we need to permute indices in the fluxes and the only available identities  are the quadratic constraints in which the fluxes are contracted with the $\kappa$ metric instead of projectors.

\end{appendix}

\end{document}